\documentclass[structabstract]{aa}
\usepackage{graphicx}
\usepackage{txfonts}
\usepackage{longtable,lscape}
\begin{document}
   \title{Rotational velocities of nearby young stars\thanks{Based upon observations with FEROS at the 2.2m MPG/ESO telescope and HARPS at the 3.6m telescope in La Silla, Chile.}}

   %\subtitle{I. Distribution of $v\sin i$}

   \author{P. Weise
          %\inst{1}
\and
R. Launhardt
\and
          J. Setiawan%\inst{2}\fnmsep\thanks{Just to show the usage          of the elements in the author field}
\and
Th. Henning
          }

\institute{Max-Planck-Institute for Astronomy (MPIA),
              K\"onigstuhl 17, 69117 Heidelberg\\
              \email{weise@mpia.de}
             }

   \date{submitted - accepted}

% \abstract{}{}{}{}{}
% 5 {} token are mandatory

  \abstract
  % context heading (optional)
  % {} leave it empty if necessary
   {Stellar rotation is a crucial parameter driving stellar magnetism, activity and mixing of chemical elements. Measuring rotational velocities of young stars can give additional insight in the initial conditions of the star formation process. Furthermore, the evolution of stellar rotation is coupled to the evolution of circumstellar disks. Disk-braking mechanisms are believed to be responsible for rotational deceleration during the accretion phase, and rotational spin-up during the contraction phase after decoupling from the disk for fast rotators arriving at the ZAMS. On the ZAMS, stars get rotationally braked by solar-type winds.}
  % aims heading (mandatory)
   {We investigate the projected rotational velocities $v\sin i$ of a sample of young stars with respect to the stellar mass and disk evolutionary state to search for possible indications of disk-braking mechanisms. Furthermore, we search for signs of rotational spin-up of stars that have already decoupled from their circumstellar disks.}
  % methods heading (mandatory)
   {We analyse the stellar spectra of 220 nearby (mostly $<100$\,pc) young (2--600\,Myr) stars for their $v\sin i$, stellar age, H$\alpha$ emission, and accretion rates. The stars have been observed with FEROS at the 2.2m MPG/ESO telescope and HARPS at the 3.6m telescope in La Silla, Chile. The spectra have been cross-correlated with appropriate theoretical templates. We build a new calibration to be able to derive $v\sin i$ values from the cross-correlated spectra. Stellar ages are estimated from the \ion{Li}{i} equivalent width at 6708\,\AA. The equivalent width and width at 10\% height of the H$\alpha$ emission are measured to identify accretors and used to estimate accretion rates $\dot{M}_\mathrm{acc}$. The $v\sin i$ is then analysed with respect to the evolutionary state of the circumstellar disks to search for indications of disk-braking mechanisms in accretors.}
  % results heading (mandatory)
   {We find that the broad $v\sin i$ distribution of our targets extends to rotation velocities of up to more than 100\,kms$^{-1}$ and peaks at a value of $7.8\pm1.2$\,kms$^{-1}$, and that $\sim70\%$ of our stars show $v\sin i<30$\,kms$^{-1}$. Furthermore, we can find indications for disk-braking in accretors and rotational spin-up of stars which are decoupled from their disks. In addition, we show that a number of young stars are suitable for precise radial-velocity measurements for planet-search surveys.}
  % conclusions heading (optional), leave it empty if necessary
   {}


   \maketitle
%
%________________________________________________________________

\section{Introduction}
\label{sec_intro}

Rotation is one of the most important kinematic properties of stars, %driving strongly the stellar evolution and 
giving rise to stellar magnetism and mixing of chemical elements. Stars form from rotating molecular cloud cores, preserving only a very minor fraction of their initial angular momentum (e.g., Palla 2002, Lamm et al. 2005). The initial angular momentum of rotating molecular cloud cores is about 4--5 orders of magnitude higher than that of the stars that eventually form in this cloud core (e.g., Bodenheimer 1989). Stellar formation models must account for that and several mechanisms are discussed, e.g., magnetic braking or magnetocentrifugally driven outflows. However, there is still no consens as on whether there is one dominant process for dispersing angular momentum during the entire star formation process or which process dominates at what evolutionary stage after the formation of a star-disk system (e.g., Palla 2002).

The mechanisms believed to be responsible for efficient loss of stellar angular momentum after the formation of a star-disk system involve transfer of angular momentum along magnetic field lines that connect the stellar surface with the disk, either onto the disk (so-called 'disk-locking', e.g. Camenzind 1990, K\"onigl 1991) or into stellar winds originating at that boundary (Matt \& Pudritz 2005 and 2008). Both models predict that the stellar magnetic field threads the circumstellar disk, accretion of disk material onto the star occurs along the field lines, and magnetic torques transfer angular momentum away from the star. In the disk-locking model, the angular velocity of the star is then locked to the Keplerian velocity at the disk boundary. For reasonable magnetic field strengths and accretion rates $\sim$\,10$^{-9}$\,M$_{\odot}$/yr, both models account for the observed rotational periods of classical T~Tauri stars (e.g., Armitage \& Clarke 1996). Both models can be observationally distinguished only by the absence or presence of stellar winds originating from the boundary. When the accretion process ends and there is no longer an efficient way to disperse angular momentum, the star can spin-up during the contraction phase (see, e.g., Lamm et al. 2005, Bouwman et al. 2006). Once arrived on the ZAMS, stars undergo additional rotational braking by magnetic winds, irrespective of whether they arrived as slow or fast rotators (e.g., Skumanich 1972, Palla 2002, Wolff et al. 2004).

If this picture holds true, very young stars are expected to rotate slower than slightly older stars that have already decoupled from their disks.
Three different classes of young stars are expected to be distinguishable, which can be interpreted as a kind of evolutionary sequence: (i) slow rotating stars which still accrete, (ii) slow to intermediate rotating stars which do not longer accrete and are about to gradually spin-up due to contraction and (iii) fast rotators without disks. However, one has to keep in mind, that all star forming regions show an enormous spread in rotation periods and that differences in rotation velocities between different star forming regions can also be caused by the initial conditions of star formation and the braking time-scale of the star-disk system (e.g., Stassun et al. 1999, Lamm et al. 2005, Herbst et al. 2007, Nguyen et al. 2009).

Several surveys of stellar rotation periods and projected rotational velocities of very young stars indeed found relatively slow rotators (e.g., Herbst et al. 2002; Rebull et al. 2004, 2006; Sicilia-Aguilar et al. 2005), which points to effective braking mechanisms, whereas other surveys did not find evidence for rotational braking of very young stars  (e.g., Makidon et al. 2004; Nguyen et al. 2009).
Nevertheless, rotational velocity measurements of stars with circumstellar disks and ongoing accretion in associations such as $\eta$\,Cha, TW~Hydrae, and NGC\,2264 support a disk-locking mechanism for the removal of angular momentum from the star (Lamm et al. 2005; Bouwman et al. 2006; Jayawardhana et al. 2006; Fallscheer \& Herbst 2006).
This rotational braking due to disk-locking is mainly visible in very young clusters with an age $<$\,3\,Myr. These stars are expected to spin-up by a factor of $\approx$\,3 due to contraction after being magnetically disconnected from the circumstellar disk. In fact, a large fraction of PMS stars have been observed that evolve at nearly constant angular velocity through the first 4\,Myr (Rebull et al. 2004).
In clusters at the age of $\sim$\,10\,Myr, signs of stellar velocity spin-up can be seen (Rebull et al. 2006).

%----
For our radial-velocity survey of very young stars, SERAM (\emph{Search for Exoplanets with Radial-velocity At the MPIA}, Setiawan et al. 2008), we started to measure the $v\sin i$ of stars with ages of 2--600\,Myr and selected slow rotators for high-precision radial velocity measurements to search for planets. These $v\sin i$ measurements are a necessary prerequisite for such an RV survey, because the projected rotational velocity of a star limits the accuracy of the radial velocity measurement due to broadening of the spectral lines. Furthermore, rotational velocities play a crucial role in describing and understanding chromospheric and coronal activity, which is related to radial velocity jitter due to stellar spots and scales with rotational velocities (e.g., Pallavicini et al. 1981, Noyes et al. 1984, Saar \& Donahue 1997). Thus, measuring rotational velocities and periods of young stars is very important to understand stellar formation, evolution and activity.

This paper is structured as follows. In section \ref{obs}, we describe the target selection, observations, and data reduction. The parameter estimation is described in Section~\ref{sec_dataana} and the calibration of $v\sin i$ is described in Appendix~\ref{sec_vm}. In Section~\ref{res}, we present the results of our analysis and search for indications of disk-braking and rotational spin-up. Finally, we summarize our results in Section \ref{con}.
%-----------------

%__________________________________________________________________

\section{Observations and data reduction}
\label{obs}
\subsection{Target list}
\label{sec_targetlist}
The target list for this survey is brightness-limited to $\mathrm{V}\leq$\,12\,mag and has been compiled from two input samples. The first list was assembled from the young star sample of the ESPRI project, the astrometric \emph{Exoplanet Search with PRIMA} (Launhardt et al. 2008), which is currently being prepared and will start in 2011. For this sub-sample, the selection criteria are young dwarf stars with spectral type F--M, age $<$\,300\,Myr, and distance $\leq$\,100\,pc. These targets have been initially selected from the FEPS data base (Carpenter et al. 2008) and the literature, e.g., Zuckermann \& Song (2004). In total, 380 young stars have been selected, which then formed the initial list for SERAM.

To this initial list, 200 stars have been added, which were selected from Torres et al. (2006). This second list consists of dwarf stars with spectral type G--M, age of $<$\,100\,Myr, and a known $v\sin i$\,$<$\,30\,km/s. Most of these additional targets are at distances $>$\,100\,pc and were therefore not included in the initial ESPRI list.

In total, our SERAM target list thus consists of 580 stars, of which 220 have been observed thus far and are presented in this paper. Of these 220 stars, 161 are from the ESPRI input list and 59 stars are from the Torres et al. (2006) list.

\subsection{Observations and data reduction}
Observations have been carried out with FEROS, the \emph{Fiber-fed Extended Range Optical Spectrograph} (Kaufer et al. 1999), at the 2.2m MPG/ESO (Max-Planck-Gesellschaft/European Southern Observatory) telescope in La Silla, Chile, within the MPG guaranteed time between November 2004 and June 2009.
With a spectral resolution of $(\lambda/\Delta\lambda)$\,=\,R\,$\approx$\,48\,000, a broad wavelength coverage of 3600--9200\,\AA\, in 39 echelle orders, and a long-term RV precision of better than 10\,$\mathrm{ms}^{-1}$ (Setiawan et al. 2007), FEROS is an appropriate instrument for planet search surveys with the radial velocity technique. The spectrograph has two fibers and can be operated both in "object-sky" (OS)-mode for spectral analysis, and in "object calibration" (OC)-mode for precise RV measurements (Kaufer et al. 2000). In the OS-mode, the first fiber is fed with the stellar spectrum, while the second fiber observes the sky background. The OC-mode is used to apply the simultaneous calibration technique, where a ThAr+Ne emission spectrum from a calibration lamp is observed simultaneously in the second fiber during the object's exposure to monitor the intrinsic velocity drift of the instrument (Baranne et al. 1996).

Both OS and OC exposure modes have been used to obtain high-resolution spectra of our target stars with signal-to-noise ratios (SNR) of 100--200 per pixel at 5500\,\AA. For a star with $m_{V}$\,=\,7\,mag, this SNR has been reached with an exposure time of 10\,min. For each star, at least one spectrum in OS-mode has been obtained. In addition, multiple epoch data has been obtained in OC mode to measure the stellar RV. However, these RV measurements are not part of this paper.\\
All data have been reduced by using the FEROS data reduction system (DRS). This package does the bias subtraction, flat-fielding, traces and extracts the echelle orders, completes and applies the wavelength calibration, and puts all data in the barycentric frame. Special care has been taken with the wavelength calibration, which was repeated if the residuals deviated by more than 1 $\sigma$ from the previous calibration.

In addition to the FEROS observations, 32 stars have been observed with HARPS (\emph{High Accuracy Radial velocity Planet Searcher}, Mayor et al. 2003) at the ESO 3.6m telescope in La Silla, Chile, in May 2008 and February 2009 (ESO periods P81 and P82). 16 of these stars have been observed with HARPS only. The other 16 stars have been observed with both FEROS and HARPS. HARPS is a fibre-fed cross-dispersed echelle spectrograph with a resolution of R\,$\approx$\,115\,000. The spectral range covered is 3780--6910\,\AA. The instrument is built to obtain very high long-term radial velocity accuracy when using the simultaneous Th-Ar mode. For the scope of this paper, the spectra have been obtained using the "HARPS\_ech\_acq\_objA" template, which acquires the stellar spectra and the CCD background. The data was reduced using the online data reduction system (DRS) at the telescope.

\section{Data analysis and parameter estimation}
\label{sec_dataana}

\subsection{Measurement of $v\sin i$}
\label{sec_vm1}
To measure the radial and rotational velocities of the target stars, the stellar spectra have been cross-correlated with appropriate numerical templates for the respective stellar spectral type. These templates have been created from synthetic spectra (see Baranne et al. 1979). For the cross-correlation, a new data analysis tool has been developed. Technically it follows Baranne et al. (1996), but a number of specific features have been added to make this tool useful for our purposes, like Gaussian fitting to multiple cross-correlation function CCF, several possibilities to measure the spectral line shape, and noise reduction filters.

For FEROS spectra, spectral lines between 3900--6800\,\AA{} are cross-correlated with the template, because the wavelength ranges $<$\,3900\,\AA{} and $>$\,6800\,\AA{} are not covered by the currently available templates. This wavelength range covers 24 echelle orders of FEROS, so that 24 separate measurements for radial and rotational velocities per star and exposure are available. The final result has been computed from the weighted mean of the 24 individual measurements. HARPS spectra have been treated in a similar way, with the difference that the whole spectrum is cross-correlated at once. A detailed technical description of this new spectra analysis tool named MACS (\emph{Max-Planck Institute for Astronomy Cross-correlation and Spectral analysis tool}) will follow in an upcoming paper (Weise et al., in preparation).

The resulting function of the cross-correlation of the stellar spectrum with the theoretical binary template is the
CCF, which is the mean profile of all involved stellar spectral lines. Thus, the minimum position of the CCF corresponds to the RV of the star, while the half width at half maximum $\sigma_{\mathrm{CCF}}$ of the CCF, derived from a Gaussian fit, corresponds to a mean spectral line width broadened due to intrinsic (thermal, natural), instrumental, and rotational effects (Benz \& Mayor 1984, Queloz et al. 1998, Melo et al. 2001):
\begin{equation}
\sigma_{\mathrm{CCF}}^{2} = \sigma_{\mathrm{rot}}^2 + \left(\sigma_{0,\star}^2 + \sigma_{\mathrm{inst}}^2\right) ,
\end{equation}
where $\sigma_{\mathrm{rot}}$ is the broadening due to stellar rotation and $\sigma_{0}^2 = \sigma_{0,\star}^2 + \sigma_{\mathrm{inst}}^2$ is the effective line width of the non-rotating star. In this latter variable, all intrinsic line-broadening effects, like thermal and gravitational broadening, turbulence (all in $\sigma_{0,\star}$), and the instrumental profile, $\sigma_{\mathrm{inst}}$, are convolved.

Thus, the projected rotational velocity, $v\sin i$, is proportional to $\sigma_\mathrm{rot}$, such that
\begin{equation}
%\label{eq_vsini1}
v\sin{i} = A \sqrt{\sigma_{\mathrm{CCF}}^{2}-\sigma_{0}^2} \, ,
\end{equation}
where $A$ is a coupling constant.

This coupling constant has already been calibrated for FEROS by Melo et al. (2001). However, it depends not only on the instrument, but also on the cross-correlation method used to calculate the CCF. Therefore, $A$ had to be recalibrated for measurements with MACS. Furthermore, $\sigma_0$ has only been calibrated for FEROS at the 1.52m telescope in La Silla, Chile. However, FEROS moved to the 2.2m telescope in 2003, such that a new calibration was needed.

Measurements of $v\sin i$  using HARPS data have often been calibrated by the calibration by Santos et al. (2002). This calibration was built from CORALIE data. Therefore, in order to have a self-consistent analysis in this paper, a new calibration similar to the one for FEROS has been calculated from our HARPS data.

The details of these calibrations are described in Appendix~\ref{sec_vm}. We derive and use in this paper: 
%\begin{equation}
%\begin{align}
$$A_\mathrm{FEROS} = 1.8\pm0.1$$
$$A_\mathrm{HARPS} = 1.88\pm0.05$$
%\end{align}
%\end{equation}
and
%\begin{equation}
%\begin{align}
$$\log\sigma_{0,\mathrm{FEROS}} = 0.641+0.043\cdot(\mathrm{b-v})$$
$$\log\sigma_{0,\mathrm{HARPS}} = 0.574+0.032\cdot(\mathrm{b-v}).$$
%\end{align}
%\end{equation}
For the 16 stars, which were observed with FEROS and HARPS, we compared the $v\sin i$ calculated by both calibrations and found that both measurements diverge no more than 0.4\,kms$^{-1}$, which is within the measurement accuracy. The accuracy for $v\sin i$ of both instruments is also similar.

\subsection{Age estimation}
\label{sec_ageest}
\begin{figure}
\centering
\includegraphics[width=9.0cm]{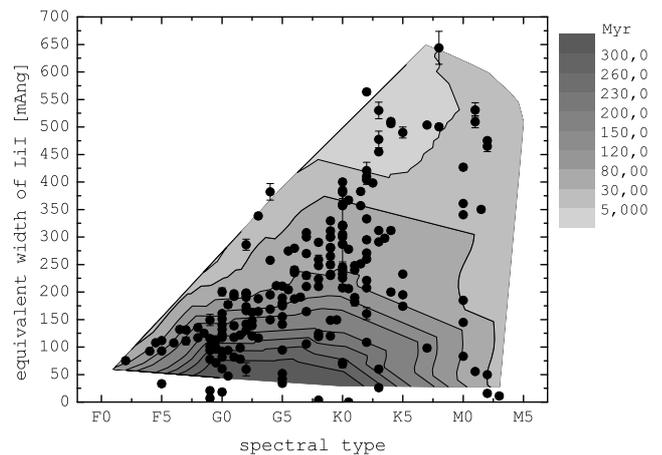}
\caption{\ion{Li}{i} measurements for our observed target stars (black dots). The shaded contours have been derived from \ion{Li}{i} measurements for young stellar associations with known mean age (Section~\ref{sec_ageest}).}
\label{fig_lithium}
\end{figure}

In the convective envelopes of F--M type stars, primordial lithium (\ion{Li}{i}) is depleted as the star ages (e.g., Palla 2002). The amount of depletion depends also on the effective temperature of the star, which means that the equivalent width (EW) of \ion{Li}{i} can only be compared among stars with the same effective temperatures, i.e., spectral type. To derive the dependency of \ion{Li}{i} EW on age and spectral type, the measured mean \ion{Li}{i} EW for different stellar associations with known age have been used. These associations are UMa (300\,Myr), Pleiades (90\,Myr), IC\,2602 (30\,Myr), TucHor (30\,Myr), $\beta$Pic (30\,Myr), TWA (8\,Myr), $\eta$Cha (8\,Myr), and NGC\,2264 (4\,Myr) and the EW of \ion{Li}{i} has been taken from Zuckerman \& Song (2004). These data points have been triangulated, interpolated and combined on a regular grid to obtain a contour plot in Figure~\ref{fig_lithium}. The effect of the rotational velocity on \ion{Li}{i} depletion is negligible for the stars in our sample (Bouvier 1995).\\
The age of the individual stars in our list has then been derived by comparing the EW of \ion{Li}{i} at 6708\,\AA{} with this grid. 
For stars that are members of young stellar associations with known age range, the age derived by the \ion{Li}{i} EW has been compared to the mean age of the associations. We find that the age estimates agree very well in all cases, confirming the validity of our approach.

\subsection{Veiling}
\label{sec_verilingm}
For accreting stars, veiling produces an additional continuum overlayed on the intrinsic stellar spectrum due to hot accretion spots on the stellar photosphere. This additional continuum alters the EW of spectral lines, as it changes the relative line depth (e.g., Appenzeller \& Mundt 1989). Therefore, the veiling has to be determined in order to derive the pure photospheric EW of \ion{Li}{i}.

In order to determine the veiling, theoretical spectra have been computed for the respective stellar spectral types by using the SPECTRUM package (Gray \& Corbally 1994) together with Kurucz atmosphere files (Kurucz 1993). These theoretical spectra have been broadened to the $v\sin i$ of the star and veiled by adding a flat continuum.
The veiling of the target stars has then been determined using spectral lines other than \ion{Li}{i} in a 50\,\AA{} window centred on the \ion{Li}{i} feature (6708\,\AA) by minimising the difference between the simulated and the observed stellar spectrum. The EW of \ion{Li}{i} was then corrected by
\begin{equation}
\label{eq_veiling}
\mathrm{EW}_\mathrm{corrected} = \mathrm{EW}_\mathrm{measured} \cdot (V+1)\,{},
\end{equation}
where $V$ is the measured veiling (Johns-Krull \& Basri 1997).

%s\subsubsection{The influence of veiling on $v\sin i$}
Veiling also affects the colours of accreting stars, such that they appear to be of earlier spectral type than they actually are. This is because the boundary layer of the circumstellar disk, where the accreted material is originating from, emits more light in the blue part of the stellar spectrum (Hartigan et al. 1989). Thus, due to an apparent shift to an earlier spectral type, this will underestimate $\sigma_0$ in Equation~(\ref{eq_sigma0}). The width of the CCF remains unchanged. With an unchanged CCF and underestimation of $\sigma_0$ due to a wrong $(\mathrm{b}-\mathrm{v})$, we therefore slightly overestimate the $v\sin i$ for accreting stars.\\
The maximum change in $v\sin i$ due to veiling can be estimated by using Equation~(\ref{eq_sigma0}) and a shift in spectral type from M0 to G0 (as an example of the effect of veiling). We found this maximum effect on $v\sin i$ to be 0.7\,kms$^{-1}$, which is smaller than the uncertainty of our $v\sin i$ measurements, and, therefore, does not affect our results.

\subsection{Signatures of accretion and disk presence}
\label{sec_halhpa_desc}

In order to analyse the measured $v\sin i$ for disk-braking mechanisms, we have to know the evolutionary state of the circumstellar disks of our target stars. Furthermore, we have to distinguish between accreting and non-accreting stars.\\
To identify stars which are still accreting, the EW and full width at 10\% height (W$_{\mathrm{H}\alpha}$) of the H$\alpha$ emission, as well as the EW of \ion{He}{i} at 5876\,\AA{} have been measured. All criteria have to be fulfilled, because high stellar rotation broadens also the H$\alpha$ emission line.\\
To distinguish between H$\alpha$ emission due to chromospheric activity and due to accretion of disk material, the criteria by Appenzeller \& Mundt (1989), and Jayawardhana et al. (2006) have both been investigated. According to Appenzeller \& Mundt (1989) a star is assumed to accrete when the EW of H$\alpha$ exceds 10\,\AA{}. In contrast, Jayawardhana et al. (2006) use W$_{\mathrm{H}\alpha}$ and determine a value of 200\,kms$^{-1}$ (4.4\,\AA{}) as the accretion threshold.

The EW of H$\alpha$ has been widely used to identify accreting stars, but chromospherically active stars can produce a comparable increase of the H$\alpha$ EW. During accretion, H$\alpha$ emission also arises from "hot spots" on the stellar surface produced by accreting disk material. The EW in both cases can be comparable, because for late-type stars, the chromospheric H$\alpha$ emission is more prominent due to a diminished photosphere as compared to hotter stars (e.g., White \& Basri 2003). This means, that the EW of H$\alpha$ due to chromospheric activity is much higher for cool stars and depends on the spectral type of the stars. Therefore, no clear statement can be made at which H$\alpha$ EW a star accretes.\\
A better discrimination between accreting and chromospherically active stars can be made by measuring the W$_{\mathrm{H}\alpha}$ of the emission feature. Emission due to an active chromosphere is limited in amount by the saturation of the chromosphere and in width by stellar rotation and non-thermal velocities of the chromosphere. White \& Basri (2003) found that accreting stars tend to have W$_{\mathrm{H}\alpha}$ broader than 270\,kms$^{-1}$ (5.9\,\AA{}), whereas Jayawardhana et al. (2006) used a less conservative threshold at 200\,kms$^{-1}$ (4.4\,\AA{}). For the purpose of this paper, the threshold from Jayawardhana et al. (2006) has been used to avoid missing accretors in this selection process.

In addition, \ion{He}{i} at 5876\,\AA{} has been investigated, because stellar rotation also affects the W$_{\mathrm{H}\alpha}$. The \ion{He}{i} feature can be directly linked to the accretion process of the star, because the luminosity of \ion{He}{i} correlates with the accretion luminosity (Fang et al. 2009 and references herein). Other features, like [\ion{O}{i}] at 8446\,\AA{} cannot be used to identify the accretion process, since it can be contaminated by the [\ion{O}{i}] feature of the circumstellar disk (Fang et al. 2009). All stars identified as accretor candidates by the H$\alpha$ criteria of Jayawardhana et al. (2006), have also been searched for presence of \ion{He}{i}.
We identified only those stars as accretors that fulfil both criteria, W$_{\mathrm{H}\alpha}>4.4\AA{}$ (Jayawardhana et al. 2006) and the presence of \ion{He}{i}. This results in 12 accretors in our sample, which are listed in Table~\ref{tab_accrstars}.\\
Accretion rates were then derived by computing the luminosity of H$\alpha$, H$\beta$, and \ion{He}{i}, which correlates with the accretion luminosity and applying Equation (2) in Fang et al. (2009).

Information about the presence or absence of circumstellar disks and the type of disks has been adopted from Carpenter et al. (2009) for those stars in our sample which have been taken from the FEPS targets.

\section{Results}
\label{res}

Our results are compiled in Table~\ref{tab_results}. In this table, we list the derived $v\sin i$ and its error in columns (3) and (4). The EW of H$\alpha$, its accuracy, W$_{\mathrm{H}\alpha}$, EW of \ion{He}{i}, and Veiling are listed in columns (5) through (9). The \ion{Li}{i} measurements and the derived stellar ages are listed in columns (10) to (13), and finally, the evolutionary state of the circumstellar disk (if present) and the stellar association to which the star belongs are listed in columns (14) and (15).

\subsection{Age}
\label{sec_ageres}
\begin{figure}
\centering
\includegraphics[width=8.0cm]{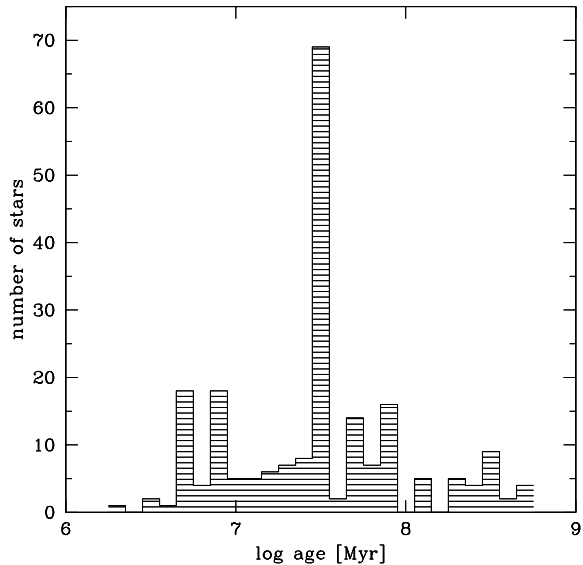}
\caption{Distribution of stellar ages within our sample, derived from the measured EW of \ion{Li}{i} at 6708 \AA{} (see Sections~\ref{sec_ageest} and~\ref{sec_ageres}).}
\label{hist_age}
\end{figure}
Almost all of our target stars show the \ion{Li}{i} feature at 6708\,\AA, thus confirming their youth. We find a typical detection limit for the \ion{Li}{i} feature of 2\,m\AA{} in spectra with a SNR$>$10 around 6700\,\AA{}. In spectra with a SNR$<$10, we were not able to measure the EW of \ion{Li}{i}. The EW of \ion{Li}{i} was not corrected for the possible blend with a nearby \ion{Fe}{i} feature, because the correction would be smaller than our measurement accuracy (cf. Soderblom 1993).\\
The measured EW of \ion{Li}{i} are listed in columns (10) and (11) of Table~\ref{tab_results} and were used to derive the age of the individual stars as described in Section~\ref{sec_ageest}. The accuracy of this method is typically 10--50\%.\\
The resulting age distribution is shown in Figure~\ref{hist_age}. The histogram has a strong peak at 30\,Myr, which is due to the selection effect towards stars in associations with these ages (from Zuckerman \& Song 2004). Note that the age estimate stated in the literature was already a selection criterion for this source sample (see Section~\ref{sec_targetlist}). In total, we found 5 stars which are most likely older than the Hyades (age $\approx$\,600\,Myr), and 43 stars that are younger than 10\,Myr.\\
In 10 stars, we could not detect the \ion{Li}{i} feature. Of these, 2 stars (TYC\,9129-1361-1 and HD\,21411) have a sufficient SNR, but the \ion{Li}{i} feature is below the detection limit of 2\,m\AA{}. The spectra of 8 stars (HD\,43989, HD\,201219, HD\,201989, HD\,205905, HD\,209253, HD209779, HD\,212291, and HD\, 225213) have an insufficient SNR\,$<$\,10, and no lithium feature could be identified and measured. These 8 stars are marked with "n/a" in Table~\ref{tab_results} (column 10).

\subsection{Evolutionary state of the stars}
\label{sec_evstate}
\begin{table*}
\caption{EW of H$\alpha$, W$_{\mathrm{H}\alpha}$, EW of \ion{He}{i}, accretion rates $\dot{M}_\mathrm{acc}$, and $v\sin i$ for accretor candidates. Close binary or multiple stars are marked with $^\dagger$.}
$$
\vspace{-1.1cm}
$$
$$
\begin{array}{lcccccl}
\hline
\hline
\\
\mathrm{Identifier} & \mathrm{EW}(\mathrm{H}\alpha) & \mathrm{W}_{\mathrm{H}\alpha} & \mathrm{EW~of~}\ion{He}{i} & \dot{M}_\mathrm{acc} & v\sin i &\mathrm{Notes}    \\
\mathrm{}           &  \mathrm{\AA}   & \mathrm{\AA} & \mathrm{\AA} & 10^{-8} M_\odot{}/\mathrm{yr} & \mathrm{kms}^{-1} &   \\
\hline
\\
\mathrm{HD}\,3221 &	2.0 &	10.2 & \ldots & \ldots	&146.6 &\mathrm{chromospheric~activity}\\
\mathrm{TWA}\,6  &	3.6 &	5.1 & \ldots & \ldots	&94.7 &\mathrm{chromospheric~activity} \\
\mathrm{V^*\,IM\,Lup} & 5.7 &	5.5 & 0.03 & \mathrm{n/a} &13.4 &\mathrm{accretor}\\
\mathrm{V^*\,HT\,Lup}^\dagger &	7.0 &	11.0 & \ldots & \ldots	&44.8 &\mathrm{chromospheric~activity}\\
\mathrm{TWA}\,5^\dagger &	8.5 & 	8.8 & \ldots & \ldots	&63.2 &\mathrm{chromospheric~activity} \\
\mathrm{CP-}72\,2713 & 12.5 & 2.3 & \ldots & \ldots &6.4 &\mathrm{chromospheric~activity}\\
\mathrm{V}\,2129\,\mathrm{Oph} &	14.6 &	6.5 & 0.25 & 1.2 &13.5 &\mathrm{accretor}\\
\mathrm{HBC\,603} & 16.1 &	7.7 & 0.1 & 0.75 &6.7 &\mathrm{accretor}\\
\mathrm{EM^{*}\,SR\,9} & 20.4 &	7.8 & 0.3 & 19.3 &14.3 &\mathrm{accretor}\\
\mathrm{V^*\,GW\,Ori}^\dagger &	20.8 &	9.9 & 0.05 & 17.1 &43.7 &\mathrm{accretor}\\
\mathrm{V\,1121\,Oph} & 20.8 &	8.3 & 0.3 & 0.5 &9.3 &\mathrm{accretor}\\
\mathrm{DoAr\,44} & 22.9 &	9.9 & 0.6 & 35.6 &15.7 &\mathrm{accretor}\\
\mathrm{V^*\,DI\,Cha} &	26.1 &	9.8 & 0.03 & 2 &38.1 &\mathrm{accretor}\\
\mathrm{V^*\,CR\,Cha} & 26.9 & 10.6 & 0.06 & 0.2 &38.5 &\mathrm{accretor}\\
\mathrm{V^*\,GQ\,Lup} &	39.5 &	13.2 & 0.6 & 13.6 &9.2 &\mathrm{accretor}\\
\mathrm{V^*\,MP\,Mus}  &	42.1 &	10.5 & 0.5 & 0.7 &13.0 &\mathrm{accretor}\\
\mathrm{V^*\,TW\,Hya} &	116.8 &	9.5 & 1.94 & 0.3 &6.2 &\mathrm{accretor}\\
\hline
\hline
\end{array}
$$
\label{tab_accrstars}
\end{table*}
In order to investigate the dependence of $v\sin i$ on the evolutionary state of the stars, we subdivided our sample into five subgroups, using the presence or absence of accretion indicators as well as the nature of the circumstellar disks (from Carpenter et al. 2009) as tracers of evolutionary state. We derive the following five subgroups: (i) stars that still accrete from their circumstellar disk (11 stars), (ii) stars with non-accreting, optically thick disks (12 stars), (iii) stars with debris disks (23 stars), (iv) stars which have no circumstellar disks left (81 stars), and (v) stars with no disk-relevant information available, but with the same mean age and age spread as group (iv). Finally, 21 stars could not be classified this way, because neither their disk status is known, nor could we derive a reliable age. These stars are not further considered in this discussion. Furthermore, we excluded those stars that are identified as SB2 binaries (see Section~\ref{sec_binary}). In Table~\ref{tab_diskbin}, we list the 5 sub-samples and their details, including the number of low-mass stars ($n_\mathrm{M-stars}$) in each sub-sample to show that no sub-sample is dominated by low-mass stars and, therefore, a low $\langle v\sin i\rangle $ is not the result of the mass-dependence of $v\sin i$ (Section~\ref{sec_fkmbin}).

All stars with EW of \ion{He}{i}\,$>$\,20\,m\AA{} (detection threshold in our spectra) \emph{and} W$_{\mathrm{H}\alpha}>$\,4.4\,\AA{} (Section~\ref{sec_halhpa_desc}) are classified here as accretors. The accretors in this sub-sample are DI\,Cha, TW\,Hya, GW\,Ori, CR\,Cha, V2129\,Oph, GQ\,Lup, IM\,Lup, DoAr\,44, EM$^*$\,SR\,9, HBC\,603, V1121\,Oph, and MP\,Mus. For our analysis, we excluded GW\,Ori because it is a close binary star (Ghez et al. 1997). The accretion rates have been calculated following Fang et al. (2009) with a typical computational accuracy of 15\%. For IM\,Lup, we were not able to derive an accretion rate because the measured EW of H$\alpha$, H$\beta$ and \ion{He}{i} could not be fitted by the assumed accretion model. In addition to the accretors, 5 more stars also show W$_{\mathrm{H}\alpha}$ over the threshold of 4.4\,\AA{}, but no \ion{He}{i} has been found in the spectra of these stars. For these stars, the H$\alpha$ emission is likely due to chromospheric activity and broadened by fast rotation or an undiscovered close binary. These 5 stars are HT\,Lup, HD\,3221, CP-72\,2713, TWA\,6, and TWA\,5. We list all 17 stars with a W$_{\mathrm{H}\alpha}>4.4$\,\AA{}, accretors and non-accretors, in Table~\ref{tab_accrstars}. 

\subsection{Projected rotational velocities}
\begin{figure}
\centering
\includegraphics[width=8.0cm]{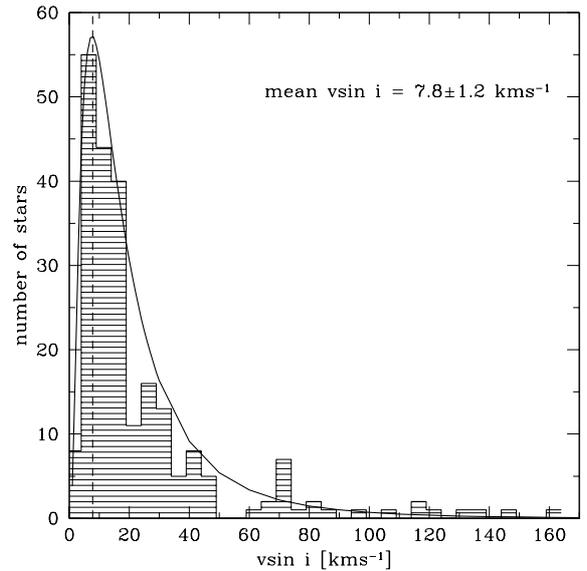}%{vsini_hist_fig5.eps}
\caption{Distribution of $v\sin i$ for our target stars. The fit of a $\log$-normal distribution function to the histogram is also shown and peaks at 7.8\,$\pm$\,1.2\,kms$^{-1}$ and 70\% of the stars have $v\sin i <$\,30\,kms$^{-1}$.}
\label{fig_vsinihist}
\end{figure}

The $v\sin i$ derived from the FEROS or HARPS spectra for all stars are listed in column (3) of Table~\ref{tab_results}. The lower limit of the $v\sin i$ measurements has been calculated by using artificially broadened theoretical spectra to be 2\,kms$^{-1}$ for our data. Thus, for all stars in our sample, the line broadening due to rotation exceeds the instrumental broadening. Figure~\ref{fig_vsinihist} shows the distribution of $v\sin i$ for all stars in our sample. We fitted all histograms with a $\log$-normal distribution function using a $\chi^2$ minimization algorithm. The fit to the $v\sin i$ distribution of all stars peaks at 7.8\,$\pm$\,1.2\,kms$^{-1}$ with a width of 9\,kms$^{-1}$. We find that 70\% of our stars have $v\sin i <$\,30\,kms$^{-1}$.
However, a sub-sample of 59 stars was biased towards slow rotators due to our selection (see Section~\ref{sec_targetlist}). 
For our analysis of $v\sin i$ as function of stellar mass (Section~\ref{sec_fkmbin}), we therefore excluded those 59 stars which have been initially chosen to have $v\sin i <$\,30\,kms$^{-1}$. However, we have to include them again in Section~\ref{sec_vsiniage} for the analysis of $v\sin i$ as a function of the evolutionary state, because these stars are the youngest in our sample. Our analysis also revealed that 8 of these 59 stars have in fact $v\sin i >$\,30\,kms$^{-1}$. These stars are CD-78~24, GW~Ori, HD~42270, CR~Cha, Di~Cha, T~Cha, HT~Lup, and HD~220054. The $v\sin i$ measurements for the stars in our sample are important to select target stars for radial velocity planet searches around young stars, since sufficient RV accuracy can only be achieved for slow rotators ($v\sin i <$\,25--30\,kms$^{-1}$). The 14 identified binary or multiple systems (see Section~\ref{sec_binary}) have also been excluded from our analysis.

\subsubsection{$v\sin i$ as a function of stellar mass}
\label{sec_fkmbin}

The rotational evolution of a star depends on its magnetic activity, which is coupled to the stellar mass (e.g., Attridge \& Herbst 1992; Palla 2002 and references herein). Lower-mass stars tend to have higher magnetic activity due to the deeper convection zones than higher-mass stars, because the stellar magnetic field is driven by the convective zone. Thus, due to the higher magnetic activity, lower-mass stars are therefore expected to rotate slower than higher-mass stars. 
To examine this dependency, we followed Nguyen et al. (2009) and divided our sample into two sub-samples of spectral types F--K (higher mass) and M (lower mass). In order to be not affected by possible age effects, we selected for this test only stars which are younger than 20\,Myr, resulting in sub-samples of 21 F--K type stars and 12 M-type stars. 
For the F--K type stars, we find a weighted mean of $\langle v\sin i\rangle =13.7\pm 0.1$\,kms$^{-1}$ and for M-type stars $\langle v\sin i\rangle =11.3\pm 0.4$\,kms$^{-1}$. This difference is small, but significant and confirms the finding of Nguyen et al. (2009), that late-type, lower-mass stars rotate on average slower than earlier-type, higher-mass stars. Note, that all stars in our sample have $M \geq 0.25 M_{\sun}$, such that we are not able to probe the higher rotation rates of stars with $M < 0.25 M_{\sun}$ as found in the Orion Nebular Cluster (Attridge \& Herbst 1992).

\subsubsection{$v\sin i$ as a function of evolutionary state}
\label{sec_vsiniage}
\label{sec_vsiniaccr}

\begin{table*}
\caption{Sub-samples used to investigate the dependence of $v\sin i$ on the evolutionary state of the stars. We list the selection criteria for the sub-sample, the number of stars and M-type lower-mass stars, mean age, $v\sin i$, and the width of the Gaussian fit to the distribution $\sigma_{v\sin i}$.}             % title of Table
\label{tab_diskbin}      % is used to refer this table in the text
\centering                          % used for centering table
\begin{tabular}{l l c ccccl}        % centered columns (4 columns)
\hline\hline                 % inserts double horizontal lines
sub-sample & description & $n_\mathrm{stars}$ & $n_\mathrm{M-stars}$ & $\langle\mathrm{Age}\rangle$ & $\langle v\sin i\rangle $ & $\sigma_{v\sin i}$ & Results\\    % table heading
 & & & & Myr & kms$^{-1}$ & kms$^{-1}$ &\\
\hline                        % inserts single horizontal line
   (i) & accreting stars & 11 & 3 & 5 & $10\pm 1$ & $3\pm 1$ & small $\langle v\sin i\rangle$, no fast rotators\\
   (ii) & non-accreting stars, thick disks & 12 & 4 & 7 & $15\pm 2$ & $8\pm 2$ & larger $\langle v\sin i\rangle$, fast rotators\\
   (iii) & stars with debris disks & 23 & 2 & 33 & $10\pm 1$ & $5\pm 1$ & small  $\langle v\sin i\rangle$, fast rotators\\
   (iv) & stars without disks & 81 & 1 & 90 & $11\pm 1$ & $4\pm 1$ & small  $\langle v\sin i\rangle$, tail of fast rotators\\
   (v) & no disk info, same age as (iv) & 81 & 3 & 90 & $7\pm 1$ & $2\pm 1$ & similar to (iv)\\
   (iv)+(v) & combined (iv)+(v) & 162 & 4 & 90 & $8\pm1$ & $2\pm1$ & small  $\langle v\sin i\rangle$, tail of fast rotators \\
\hline                                   %inserts single line
\end{tabular}
\end{table*}

\begin{figure*}
\centering
\includegraphics[width=8.0cm]{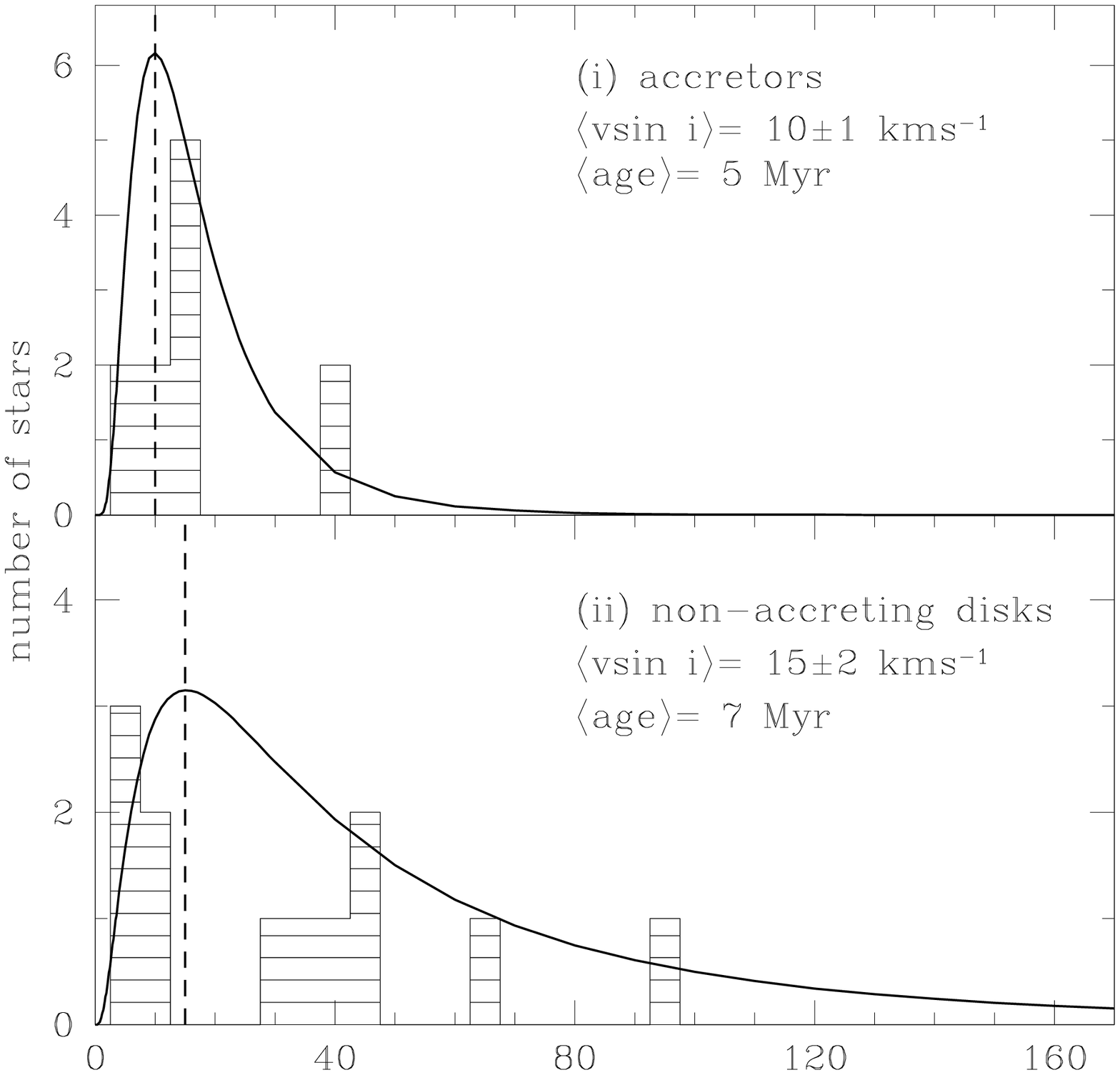}%{fig_diskbin_a_2.eps}
\includegraphics[width=8.0cm]{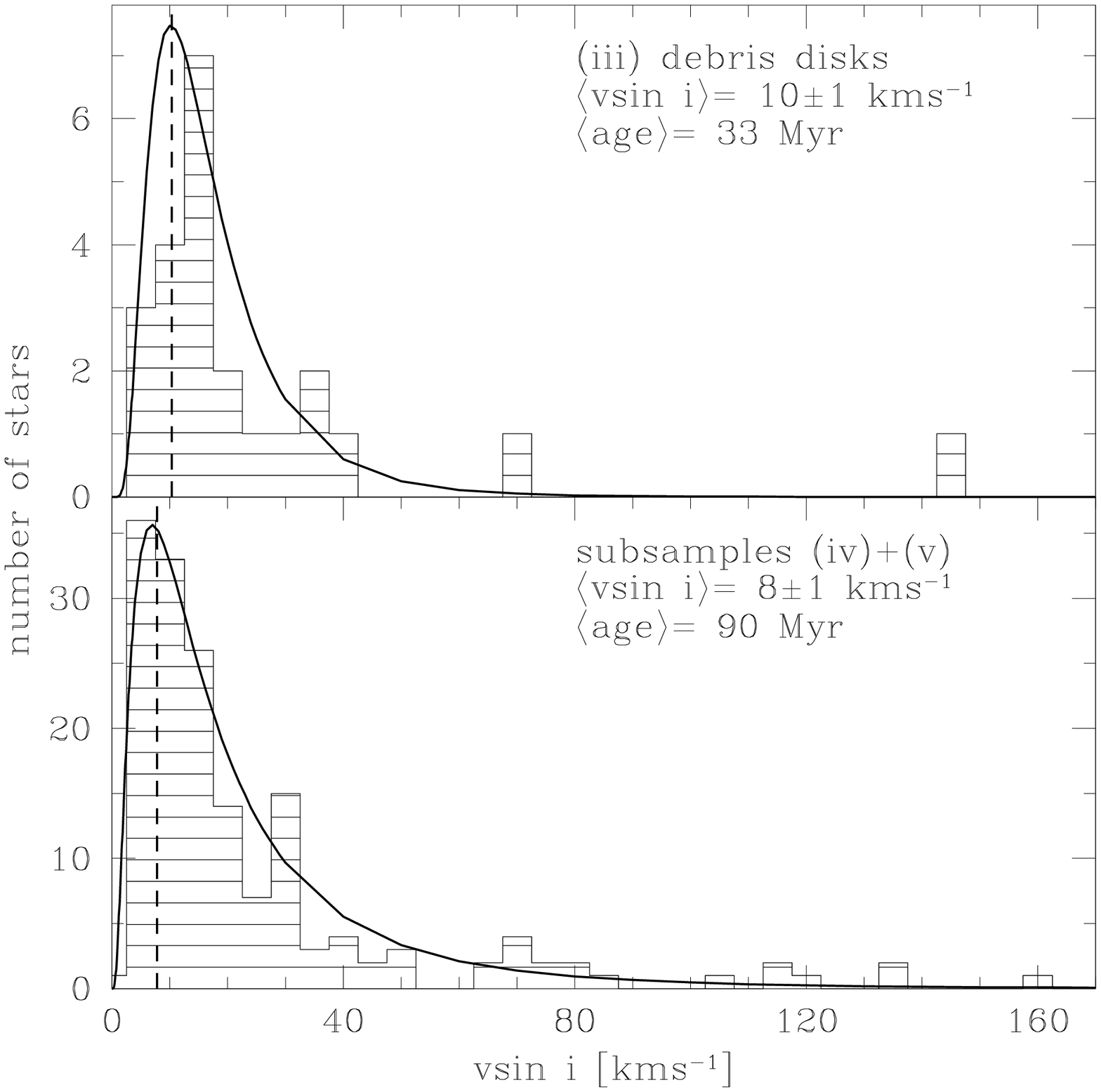}%{fig_diskbin_b_2.eps}
\caption{Distribution of $v\sin i$ for the 4 sub-samples of stars, (i) accretors, (ii) stars with non-accreting, thick disks, (iii) stars with debris disks, (iv)+(v) stars without disks and stars without information of circumstellar disks. The mean $v\sin i$ of the $\log$-normal distribution function fits to the histograms are shown by dashed lines.}
\label{fig_vsini_disks}
\end{figure*}
To search for evidence of disk-braking and rotational spin-up in our sample, we analysed the $v\sin i$ distributions of the 5 sub-samples described in Section~\ref{sec_evstate}. Note that these sub-samples have been created independently of the mass of the individual stars and we only compare different evolutionary stages. The $v\sin i$ distributions of the sub-samples have been fitted with a $\log$-normal distribution function. The resulting mean $\langle v\sin i\rangle $ and widths $\sigma_{v\sin i}$ are listed in columns 6 and 7 of Table~\ref{tab_diskbin}. Since sub-samples (iv) and (v) represent similar evolutionary states, we combined them for further analysis. The distribution histograms together with the fits are shown in Figure~\ref{fig_vsini_disks}.

As a result, we find that accretors are slow rotators. Furthermore, stars in sub-sample (ii) with non-accreting and optically thick disks rotate on average faster compared to accreting stars and stars at later evolutionary stages. However, the $v\sin i$ distribution for these stars shows a large spread of rotation velocities. Stars older than 30\,Myr rotate again slower on average, but with a long tail in the distribution towards individual fast rotators.  This general trend conforms with the well-known picture that rotational spin-up of contracting Pre-Main-Sequence stars is compensated by efficient braking mechanisms during the accretion phase. At the end of the accretion phase, there is no or less efficient braking, such that a rotational spin-up occurs. In later evolutionary stages on the ZAMS, other braking mechanisms are efficient. These mechanisms depend on different disk-braking and contraction times of the individual stars, thus leading to several individual fast rotators. 
Note that other surveys of stellar rotation in young open clusters found a larger spread in rotation of stars with an age of 30\,Myr than we found in our data and rotational spin-down has been found for stars older than 30\,Myr (e.g., Allain 1998). This difference is possibly due to a lower number of stars in sub-sample (iii) and the error in age estimation. However, the general trend in our data is consistent with the findings of other surveys (e.g., Scilia-Aguilar et al. 2004; Lamm et al. 2005).

To search for indications of disk-braking, we compared the $v\sin i$ of stars in sub-sample (i) with the accretion rates listed in Table~\ref{tab_accrstars}. Disk-braking mechanisms depend on a magnetic connection between the star and its circumstellar disk and therefore on the accretion rate of the star. Furthermore, stellar age and disk lifetimes are expected to have an impact on disk-braking, since the amount of time spent for rotational braking by the disk (disk-braking time; e.g., Nguyen et al. 2009) determines the resulting stellar rotation velocity. However, this also depends on the initial rotational conditions of the star-forming region. The $v\sin i$ distribution for accretors in sub-sample (i) can be due to a combination of these effects, in which the amount of time spent for disk-braking is dominant, because the two youngest accretors (DI\,Cha and CR\,Cha) rotate faster than the other accretors independent of their accretion rate. 
This gets also supported by other surveys (e.g., Joergens \& Guenther 2001; Scilia-Aguilar et al. 2004; Lamm et al. 2005; Jayawardhana et al. 2006; Nguyen et al. 2009). 
A larger survey of stars in different star-forming regions, measuring the age and accretion rates, would be necessary to study the dependence of stellar rotation on $\dot{M}_\mathrm{acc}$ further.

\subsection{Stellar rotation periods}
\begin{figure}
\centering
\includegraphics[width=8.0cm]{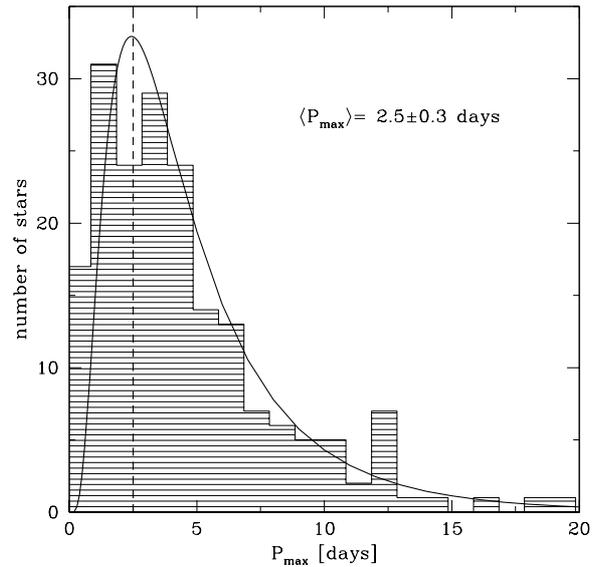}
\caption{Distribution of the stellar rotation period $P_{\mathrm{max}}$ for the whole sample. The mean $\langle{}P_\mathrm{max}\rangle = 2.5\pm0.3$\,days, derived by fitting a $\log$-normal distribution function, is marked by the dashed line.}
\label{fig_psini}
\end{figure}

For stars for which accurate measurements of T$_\mathrm{eff}$, absolute magnitude, and bolometric correction are available (e.g., from Weise 2007; Flower 1996), we estimated the stellar radius $R/R_{\odot}$ following Valenti \& Fischer (2005).
An upper limit to the rotational period of the star can then be derived from the estimated radius by using:
\begin{equation}
P_{\mathrm{max}} = 50.6\,\frac{R/R_{\odot}}{v\sin i} \qquad [d].
\end{equation}

The resulting distribution of rotational periods for the whole sample is shown in Figure~\ref{fig_psini}. The distribution of $P_{\mathrm{max}}$ for all our target stars peaks at at $\langle P_{\mathrm{max}}\rangle = 2.5\pm$0.3\,days, derived by fitting a $\log$-normal distribution function. This compares well to the result derived by Cieza \& Baliber (2007), that young stars show rotational periods around 2--3\,days.\\
However, because of the uncertainties introduced by the assumptions used to calculate the radii of the stars, no further analysis of $P_\mathrm{max}$ has been done, like, e.g. for $v\sin i$ in Section~\ref{sec_vsiniage}.\\

\subsection{Binary and multiple systems}
\label{sec_binary}
From the CCFs, we are also able to identify double-lined spectroscopic binaries (SB2). We clearly identified 9 stars to be SB2s (GJ\,3323, TWA\,4, TYC\,6604-118-1, TYC\,9129-1361-1, HD\,124784, HD\,139498, HD\,140374, HD\,141521, and HD\,155555), such that our total list in Table~\ref{tab_results} consists of 229 stars. All 9 stars have already been known as multiple systems. For all 9 SB2s, the CCF of each component has been analysed separately and the appropriate $v\sin i$ has been calculated, assuming that both stars have the same color index. This assumption holds approximately true, when both CCFs of the systen are of similar depth, which is the case for these 9 stars. In Table~\ref{tab_results}, the secondary components are marked by the main identifier and "B", but we did not derive the age.

In addition to the clearly identified SB2 stars, more known multiple stars are in our sample: TWA\,5, GW\,Ori, DI\,Cha, and HT\,Lup (Jayawardhana 2006; Ghez et al. 1997; Mathieu et al. 1995). However, the components of these systems cannot be distinguished in the CCFs to analyse them separately, because they are single-lined binaries (SB1). A few more SB1 systems might be found in our sample when the analysis of multiple epoch RV data is finished, but this is beyond the scope of this paper.

\addtocounter{table}{1}

\section{Summary \& Conclusions}
\label{con}

We analysed 229 young and nearby stars for their projected rotational velocity $v\sin i$, stellar age, and accretion signatures. The stars were part of our initial RV planet search target list in the SERAM project and have been observed with FEROS at the 2.2m MPG/ESO telescope and HARPS at the 3.6m telescope in La Silla, Chile. For stars showing broad H$\alpha$ emission, we checked other accretion indicators, like the EW of \ion{He}{i} at 5876\,\AA, to identify those stars that still accrete material from a circumstellar disk. For these stars, the veiling has also been measured. The age of the stars has been derived from the EW of \ion{Li}{i} at 6708\,\AA{}.

We calculated $v\sin i$ of our 220 target stars and 9 spectroscopic companions by measuring the width of the CCF of the stellar spectra. To do this, we cross-correlated the stellar spectra with theoretical templates of similar spectral type. The CCF has been fitted with a Gaussian and the conversion to $v\sin i$ was done with our new calibration (Equation~\ref{eq_sigma0}) for both FEROS and HARPS spectra.

The main conclusions from this work can be summarized as follows:
\begin{itemize}
%\item{}
\item{The mean stellar age of our sample is 30\,Myr with a spread (width of Gaussian fit) of 20\,Myr. 43 stars are younger than 10\,Myr.}
\item{The distribution of $v\sin i$ for all target stars peaks at 7.8$\pm$1.2\,kms$^{-1}$, with the vast majority having projected rotational velocities $<30$\,kms$^{-1}$.}
\item{We found indication for rotational braking due to disk-locking, because the accreting stars in our sample rotate significantly slower ($\langle v\sin i\rangle =10\pm1$\,kms$^{-1}$) than the non-accreting stars with more evolved disks ($\langle v\sin i\rangle =15\pm2$\,kms$^{-1}$). The only 2 fast rotating ($v\sin i \sim 38$\,kms$^{-1}$) accretors in our sample are DI\,Cha and CR\,Cha. This might point to differences in time spent for disk-braking or to different initial conditions in different star formation regions, but the significance of this statement is hampered by low number statistics.}
\item{This $v\sin i$ evolution also provides indication that stars undergo rotational spin-up after being decoupled from their circumstellar disk and evolving along the Pre-Main-Sequence track. For more evolved stars with debris or no disks and an age\,$>30$\,Myr, $\langle v\sin i\rangle$ decreases again to 10\,$\pm$\,1\,kms$^{-1}$. We conclude that these stars have been rotationally braked on the ZAMS, e.g., by magnetic winds.
}
\item{We estimated the maximum rotational period of our target stars. The period distribution of the whole sample peaks at 2.5\,$\pm$\,0.3\,days, a similar result as found for young stars by Cieza \& Biliber (2007).}
\item{Due to disk-braking, young and accreting stars can rotate sufficiently slow to serve as suitable targets for RV planet searches, since the accuracy of radial velocity measurements depends on the stellar $v\sin i$. When accretion ends after $\sim$5--10\,Myr and the stars decouple from their disk, they tend to spin up as they contract, such that they are usually no longer suitable for precise RV measurements. This changes only when the stars arrive on the ZAMS and are slowed down again by wind braking at ages of $\sim$30\,Myr.}
\end{itemize}

\begin{acknowledgements}
This research has made use of the SIMBAD database, operated at CDS, Strasbourg, France. Based on observations made with the European Southern Observatory telescopes obtained from the ESO/ST-ECF Science Archive Facility. We gratefully thank M. Fang for the calculation of the accretion rates, R. Mundt for helpful discussions, the unknown Referee for his comments, and the staff at La Silla, Chile for their support. We want to thank the observers E. Meyer, M. Zechmeister, M. Lendl, V. Joergens, J. Datson, T. Schulze-Hartung, D. Fedele, J. Carson and A. M\"uller.
\end{acknowledgements}

{\scriptsize
\longtabL{3}{
\renewcommand{\footnoterule}{}
\begin{landscape}
\begin{longtable}{llrrrrrcrrrrrll}
\caption{Table of results. In Column (1) star identifier, column (2) the spectral type (from SIMBAD), the other columns are described in Section~\ref{res}. The EW of H$\alpha$ (col. 5) is negative when in emission, "a" for absorption, and "fi" for filled up H$\alpha$. The evolutionary state of the circumstellar disk (col. 14) is represented by "accr." for accretion disk, "non-accr." for non-accreting and optically thick disks, and "debris" for disks with cold dust only, "none" for verified absence of a circumstellar disk. No entry means no data is available. Stars marked with $^\dagger$ have been taken from the Torres et al. (2006) list.\\
References: $^a$:\,Brown et al. (2007), $^b$:\,Carpenter et al. (2009), $^c$:\,Chen et al. (2005), $^d$:\,Gautier et al. (2008), $^e$:\,Kessler-Silacci et al. (2006), $^f$:\,K\"ohler (2001), $^g$:\,Lommen et al. (2007), $^h$:\,L\'opez-Santiago et al. (2006), $^i$:\,Low et al. (2005), $^k$:\,Makarov \& Fabricius (2001), $^l$:\,Mamajek et al. (2002), $^m$:\,Mamajek et al. (2004), $^n$:\,Neuh\"auser et al. (2003), $^o$:\,Padgett et al. (2006), $^p$:\,Rebull et al. (2008), $^r$:\,Soderblom et al. (1998), $^s$:\,Torres et al. (2000), $^t$:\,Wichmann et al. (2003), $^u$:\,Zuckerman \& Song (2004), $^v$:\,SIMBAD, $^w$:\,Weise (2007)}\label{tab_results}\\
\hline\hline
(1) & (2) & (3) & (4) & (5) & (6) & (7) & (8) & (9) & (10) & (11) & (12) & (13) & (14) & (15) \\
Star ID   & SpT$^{v,w}$ & $v\sin i$ & $\sigma_{v\sin i}$ & EW(H$\alpha$) & $\sigma_{\mathrm{EW(H}\alpha)}$ & W$_{\mathrm{H}\alpha}$ & EW(\ion{He}{i}) & Veiling & EW(\ion{Li}{i}) & $\sigma_{\mathrm{EW(\ion{Li}{i})}}$ & Age & $\sigma_{\mathrm{Age}}$ & disk$^b$ & YSA$^v$ \\
%\hline
& & kms$^{-1}$ & kms$^{-1}$ & \AA & \AA & \AA & \AA & & m\AA & m\AA & Myr & Myr & & \\
%\hline
\hline
\endfirsthead
\caption{continued.}\\
\hline\hline
(1) & (2) & (3) & (4) & (5) & (6) & (7) & (8) & (9) & (10) & (11) & (12) & (13) & (14) & (15) \\
Star ID   & SpT$^{v,w}$ & $v\sin i$ & $\sigma_{v\sin i}$ & EW(H$\alpha$) & $\sigma_{EW(\mathrm{H}\alpha)}$ & W$_{\mathrm{H}\alpha}$ & EW(\ion{He}{i}) & Veiling & EW(\ion{Li}{i}) & $\sigma_{\mathrm{EW(\ion{Li}{i})}}$ & Age & $\sigma_{\mathrm{Age}}$ & disk$^b$ & YSA$^v$ \\
%\hline
& & kms$^{-1}$ & kms$^{-1}$ & \AA & \AA & \AA & \AA & & m\AA & m\AA & Myr & Myr & & \\
%\hline
\hline
\endhead
\hline
\endfoot
\object{HD 105}  & G1 & 14.7 & 1.0 & a & \ldots & \ldots & \ldots & \ldots & 145 & 4 & 50 & 15 & debris & Tuc/Hor$^u$ \\
\object{HD 225213}  & M1.5 & 5.8 & 1.7 & a & \ldots & \ldots & \ldots & \ldots & n/a & \ldots & \ldots & \ldots & \ldots & field \\
\object{HD 377}  & G6 & 14.2 & 1.0 &  & \ldots & \ldots & \ldots & \ldots & 131 & 10 & 90 & 15 & debris & field$^b$ \\
\object{HD 987}  & G5 & 7.3 & 1.1 & a & \ldots & \ldots & \ldots & \ldots & 191 & 5 & 30 & 15 & none$^p$ & Tuc/Hor$^u$ \\
\object{HD 984}  & F6 & 41.8 & 5.1 & a & \ldots & \ldots & \ldots & \ldots & 131 & 5 & 25 & 10 & none & field$^b$ \\
\object{HD 1466}  & F8 & 22.7 & 2.2 & a & \ldots & \ldots & \ldots & \ldots & 117 & 3 & 30 & 10 & debris$^m$ & Tuc/Hor$^u$ \\
\object{HD 3221}  & K4.5 & 146.6 & 11.5 & -2.03 & 0.05 & 10.2 & \ldots & \ldots & 174 & 5 & 30 & 10 & debris$^m$ & Tuc/Hor$^u$ \\
\object{CD-78 24}$^\dagger$ & K3 & 30.7 & 2.3 & -0.1 & 0.01 & 1.5 & \ldots & \ldots & 291 & 8 & 15 & 10 & \ldots & field \\
\object{HD 4944}  & F8.5 & 33.5 & 1.9 & a & \ldots & \ldots & \ldots & \ldots & 132 & 5 & 30 & 10 & \ldots & field$^t$ \\
\object{TYC 8852-264-1 } & G9 & 31.7 & 1.8 & -0.21 & 0.03 & 2.4 & \ldots & \ldots & 287 & 5 & 20 & 10 & none & Tuc/Hor$^u$ \\
\object{HD 7661}  & G7.5 & 3.8 & 1.9 & a & \ldots & \ldots & \ldots & \ldots & 34 & 5 & 300 & 50 & none & field$^b$ \\
\object{HIP 6276}  & G6 & 4.2 & 1.7 & a & \ldots & \ldots & \ldots & \ldots & 141 & 5 & 50 & 10 & debris & AB\,Dor$^u$ \\
\object{HD 8813}  & G3.5 & 6.7 & 1.2 & a & \ldots & \ldots & \ldots & \ldots & 120 & 4 & 120 & 20 & \ldots & field$^t$ \\
\object{HD 8558}  & G5.5 & 13.6 & 0.9 & a & \ldots & \ldots & \ldots & \ldots & 189 & 3 & 30 & 15 & debris$^m$ & Tuc/Hor$^u$ \\
\object{CD-31 571}$^\dagger$ & K2 & 22.3 & 1.5 & fi & \ldots & \ldots & \ldots & \ldots & 260 & 5 & 30 & 10 & \ldots & field \\
\object{HD 9054}  & K1 & 3.9 & 1.9 & a & \ldots & \ldots & \ldots & \ldots & 182 & 7 & 30 & 10 & \ldots & Tuc/Hor$^u$ \\
\object{TYC 8047-232-1}  & K2.5 & 20.6 & 1.2 & -0.11 & 0.01 & 1.8 & \ldots & \ldots & 333 & 5 & 30 & 10 & \ldots & Tuc/Hor$^u$ \\
\object{HD 12039}  & G2.5 & 15.8 & 1.0 & a & \ldots & \ldots & \ldots & \ldots & 173 & 4 & 30 & 15 & debris & Tuc/Hor$^u$ \\
\object{1RXS J020154.6-523453}  & K3.5 & 19.9 & 1.2 & -0.18 & 0.02 & 2.0 & \ldots & \ldots & 298 & 5 & 20 & 10 & \ldots & Tuc/Hor$^u$ \\
\object{HD 13246}  & F7.5 & 43.3 & 2.7 & a & \ldots & \ldots & \ldots & \ldots & 125 & 5 & 30 & 15 & \ldots & Tuc/Hor$^u$ \\
\object{HD 14706}  & G0 & 7.9 & 1.0 & a & \ldots & \ldots & \ldots & \ldots & 99 & 5 & 90 & 10 & \ldots & field \\
\object{HD 15526}  & G4 & 8.9 & 1.0 & a & \ldots & \ldots & \ldots & \ldots & 155 & 3 & 60 & 15 & none & field$^b$ \\
\object{GSC 08056-00482}  & M2.5 & 29.3 & 1.7 & -4.98 & 0.6 & 3.5 & \ldots & \ldots & 350 & 5 & 20 & 10 & none$^p$ & Tuc/Hor$^u$ \\
\object{V* AF Hor} & M2.5 & 19.0 & 1.3 & -4.07 & 0.2 & 2.1 & \ldots & \ldots & 16 & 5 & 20 & 10 & debris$^m$ & Tuc/Hor$^u$ \\
\object{HIP 12545}  & K9.5 & 9.0 & 1.0 & -1.07 & 0.05 & 3.0 & \ldots & \ldots & 427 & 5 & 10 & 3 & none$^p$ & $\beta$\,Pic$^u$ \\
\object{TYC 8497-995-1}  & K9 & 5.9 & 1.4 & -0.64 & 0.05 & 2.0 & \ldots & \ldots & 98 & 5 & 30 & 10 & \ldots & Tuc/Hor$^u$                   \\
\object{HD 17662}$^\dagger$ & G5 & 17.0 & 1.2 & a & \ldots & \ldots & \ldots & \ldots & 42 & 2 & 250 & 50 & \ldots & field \\
\object{HD 17925}  & K0 & 4.8 & 1.6 & a & \ldots & \ldots & \ldots & \ldots & 206 & 5 & 90 & 10 & \ldots & field$^b$ \\
\object{CD-37 1123}$^\dagger$ & G9 & 6.9 & 1.2 & a & \ldots & \ldots & \ldots & \ldots & 230 & 5 & 30 & 15 & \ldots & field \\
\object{HD 19330}  & F9 & 7.0 & 1.1 & a & \ldots & \ldots & \ldots & \ldots & 18 & 5 & 400 & 50 & \ldots & \ldots \\
\object{HD 19491} & G1 & 5.2 & 1.5 & a & \ldots & \ldots & \ldots & \ldots & 33 & 5 & 600 & 50 & \ldots & \ldots \\
\object{HD 19668}  & G5.5 & 7.1 & 1.1 &  & \ldots & \ldots & \ldots & \ldots & 165 & 5 & 90 & 10 & debris & field$^b$ \\
\object{HD 21411}  & G6 & 4.1 & 1.8 & a & \ldots & \ldots & \ldots & \ldots & $<$2 & \ldots & $>$600 & \ldots & none & field$^b$ \\
\object{TYC 5879-490-1}  & G4 & 14.7 & 1.0 & a & \ldots & \ldots & \ldots & \ldots & 188 & 4 & 30 & 15 & none & field$^b$ \\
\object{RX J 0329.1+0118}  & F9 & 65.1 & 4.0 & a & \ldots & \ldots & \ldots & \ldots & 82 & 5 & 200 & 50 & none & field$^b$ \\
\object{1RXS J033149.8-633155}$^\dagger$ & G9 & 18.1 & 1.1 & a & \ldots & \ldots & \ldots & \ldots & 300 & 3 & 25 & 15 & \ldots & field \\
\object{HD 21955}$^\dagger$ & G7 & 27.7 & 1.7 & a & \ldots & \ldots & \ldots & \ldots & 230 & 1 & 30 & 10 & \ldots & field \\
\object{HD 23208}$^\dagger$ & G8 & 9.3 & 0.4 & a & \ldots & \ldots & \ldots & \ldots & 252 & 3 & 25 & 10 & \ldots & \ldots \\
\object{RX J 0354.4+0535}  & G2 & 69.6 & 5.6 & a & \ldots & \ldots & \ldots & \ldots & 118 & 5 & 90 & 10 & none & field$^b$ \\
\object{HD 25300}  & K4 & 13.0 & 1.0 & -0.62 & 0.05 & 4.0 & \ldots & \ldots & 108 & 5 & 90 & 10 & none & field$^b$ \\
\object{TYC 5882-1169-1}  & K2 & 6.5 & 1.3 & a & \ldots & \ldots & \ldots & \ldots & 221 & 5 & 30 & 15 & \ldots & Tuc/Hor$^u$ \\
\object{HD 25457}$^\dagger$ & F6.5 & 18.5 & 1.1 & a & \ldots & \ldots & \ldots & \ldots & 112 & 3 & 50 & 15 & \ldots & AB\,Dor$^u$ \\
\object{CD-31 1688}$^\dagger$ & G6 & 30.0 & 1.8 & a & \ldots & \ldots & \ldots & \ldots & 240 & 4 & 30 & 10 & \ldots & field \\
\object{HD 26990}  & G2 & 7.1 & 1.1 & a & \ldots & \ldots & \ldots & \ldots & 47 & $<$1 & 300 & 50 & none & field$^r$ \\
\object{CD-43 1395}$^\dagger$ & G7 & 28.1 & 1.7 & a & \ldots & \ldots & \ldots & \ldots & 270 & 4 & 30 & 10 & \ldots & field \\
\object{CD-43 1451}$^\dagger$ & G9 & 21.3 & 1.4 & fi & \ldots & \ldots & \ldots & \ldots & 280 & 4 & 15 & 5 & \ldots & field \\
\object{TYC 5891-69-1}$^\dagger$ & G4 & 20.5 & 1.3 & fi & \ldots & \ldots & \ldots & \ldots & 280 & 4 & 10 & 5 & none & field \\
\object{1RXS J043451.0-354715}$^\dagger$ & K1 & 11.3 & 1.0 & -0.24 & 0.03 & 1.9 & \ldots & \ldots & 300 & 5 & 20 & 15 & \ldots & field \\
\object{RX J 0434.3+0226}  & K5.5 & 5.7 & 1.4 & -0.67 & 0.03 & 2.2 & \ldots & \ldots & 233 & 3 & 30 & 10 & none & field$^b$ \\
\object{HD 29623}  & G2 & 8.8 & 1.0 & a & \ldots & \ldots & \ldots & \ldots & 96 & 1 & 90 & 10 & \ldots & field$^t$ \\
\object{RX J 0442.5+0906}  & K0 & 12.1 & 0.9 & a & \ldots & \ldots & \ldots & \ldots & 240 & 5 & 60 & 15 & none & field$^b$ \\
\object{HIP 23309}  & K7 & 5.8 & 1.5 & -1.21 & 0.03 & 2.3 & \ldots & \ldots & 341 & 3 & 10 & 3 & none$^p$ & $\beta$\,Pic$^u$ \\
\object{HD 31950}$^\dagger$ & G & 29.2 & 0.8 & a & \ldots & \ldots & \ldots & \ldots & 181 & 6 & 30 & 15 & none$^b$ & field \\
\object{GJ 3323}  & M1.5 & 10.1 & 3.8 & -5.43 & 0.2 & 2.7 & \ldots & \ldots & 11 & 5 & 12 & 15 & \ldots & $\beta$\,Pic$^u$ \\
\object{GJ 3323B}  & \ldots & 11.3 & 1.5 & \ldots & \ldots & \ldots & \ldots & \ldots & \ldots & \ldots & \ldots & \ldots & \ldots & \ldots \\
\object{HD 35114}  & F7 & 73.8 & 5.7 & a & \ldots & \ldots & \ldots & \ldots & 108 & 1 & 30 & 10 & \ldots & Tuc/Hor$^u$ \\
\object{HD 35850}  & G0 & 71.6 & 9.6 & a & \ldots & \ldots & \ldots & \ldots & 149 & 10 & 12 & 15 & \ldots & $\beta$\,Pic$^u$ \\
\object{V* AB Dor} & K1 & 104.5 & 10.9 & a & \ldots & \ldots & \ldots & \ldots & 233 & 19 & 50 & 15 & none$^c$ & AB\,Dor$^u$ \\
\object{V* GW Ori}$^\dagger$ & K3 & 43.7 & 2.4 & -20.76 & 0.2 & 9.9 & 0.05 & 0.95 & 240 & 5 & 5 & 2 & accr.$^e$ & Ori \\
\object{TYC 7600-0516-1}  & K0.5 & 18.6 & 1.1 & a & \ldots & \ldots & \ldots & \ldots & 248 & 5 & 30 & 15 & \ldots & Tuc/Hor$^u$ \\
\object{HD 37484}  & F3 & 48.4 & 4.0 & a & \ldots & \ldots & \ldots & \ldots & 92 & 3 & 30 & 15 & \ldots & field$^b$ \\
\object{V* AI Lep} & G0 & 27.9 & 1.6 & a & \ldots & \ldots & \ldots & \ldots & 200 & 7 & 30 & 15 & \ldots & field$^b$ \\
\object{HD 38397}  & G1 & 15.9 & 1.0 & a & \ldots & \ldots & \ldots & \ldots & 151 & 4 & 60 & 15 & \ldots & field$^t$ \\
\object{HD 38207}  & F2 & 70.7 & 4.2 & a & \ldots & \ldots & \ldots & \ldots & 75 & 5 & 55 & 15 & \ldots & field$^b$ \\
\object{HD 38949}  & G1 & 7.2 & 1.1 & a & \ldots & \ldots & \ldots & \ldots & 82 & $<$1 & 250 & 50 & \ldots & field$^b$ \\
\object{HD 42270}$^\dagger$ & K0 & 32.6 & 2.3 & a & \ldots & \ldots & \ldots & \ldots & 305 & 1 & 20 & 10 & \ldots & field \\
\object{HD 40216}  & F7 & 32.2 & 1.8 & a & \ldots & \ldots & \ldots & \ldots & 115 & 2 & 30 & 10 & \ldots & Tuc/Hor$^u$ \\
\object{HD 41700}  & F8.5 & 16.6 & 1.0 & a & \ldots & \ldots & \ldots & \ldots & 78 & 5 & 200 & 50 & \ldots & field$^r$ \\
\object{CD-34 2676}$^\dagger$ & K1 & 18.1 & 1.1 & \ldots & \ldots & \ldots & \ldots & \ldots & 240 & 5 & 65 & 15 & \ldots & field \\
\object{HD 45081}  & K3.5 & 15.9 & 1.0 & -0.79 & 0.02 & 2.1 & \ldots & \ldots & 398 & 3 & 10 & 5 & none & $\beta$\,Pic$^u$ \\
\object{V* AB Pic} & K0.5 & 11.5 & 0.9 & a & \ldots & \ldots & \ldots & \ldots & 287 & 4 & 30 & 15 & none & Tuc/Hor$^u$ \\
\object{HD 43989}  & G0.5 & 42.3 & 5.0 & \ldots & \ldots & \ldots & \ldots & \ldots & n/a & \ldots & \ldots & \ldots & debris & Tuc/Hor$^u$ \\
\object{HD 45270}  & G0 & 17.2 & 1.1 & a & \ldots & \ldots & \ldots & \ldots & 139 & 5 & 50 & 10 & \ldots & AB\,Dor$^u$ \\
\object{HD 47875}  & G3 & 11.6 & 0.9 & a & \ldots & \ldots & \ldots & \ldots & 198 & 3 & 30 & 15 & none & field$^b$ \\
\object{HD 48189}  & G0 & 16.5 & 1.0 & a & \ldots & \ldots & \ldots & \ldots & 132 & 3 & 50 & 10 & none$^c$ & AB\,Dor$^u$ \\
\object{HD 51062}$^\dagger$ & G5 & 14.2 & 0.9 & a & \ldots & \ldots & \ldots & \ldots & 180 & 2 & 200 & 50 & \ldots & Carina-Vela \\
\object{TYC 8545-1758-1}  & F9 & 46.7 & 6.8 & a & \ldots & \ldots & \ldots & \ldots & 136 & 5 & 25 & 10 & \ldots & field$^b$ \\
\object{HD 51797}$^\dagger$ & K0 & 14.5 & 0.5 & a & \ldots & \ldots & \ldots & \ldots & 288 & 5 & 30 & 15 & \ldots & \ldots \\
\object{HD 55279}  & K2 & 9.4 & 1.0 & a & \ldots & \ldots & \ldots & \ldots & 271 & 4 & 30 & 15 & \ldots & Tuc/Hor$^u$ \\
\object{CD-84 80}$^\dagger$ & G9 & 7.7 & 1.1 & a & \ldots & \ldots & \ldots & \ldots & 300 & 5 & 30 & 15 & \ldots & field \\
\object{HD 61005}  & G5.5 & 9.9 & 0.9 & a & \ldots & \ldots & \ldots & \ldots & 169 & 3 & 30 & 15 & debris & field$^r$ \\
\object{HD 62850}  & F9.5 & 14.4 & 1.0 & a & \ldots & \ldots & \ldots & \ldots & 122 & 1 & 90 & 10 & \ldots & field$^r$ \\
\object{HD 70573}  & G2 & 13.5 & 0.9 & a & \ldots & \ldots & \ldots & \ldots & 148 & 4 & 70 & 15 & \ldots & field$^b$ \\
\object{V* EG Cha}$^\dagger$ & K4 & 21.4 & 1.3 & -0.96 & 0.05 & 2.9 & \ldots & \ldots & 510 & 6 & 5 & 2 & none$^d$ & Cha \\
\object{RX J 0849.2-7735}  & K2 & 6.0 & 1.4 & a & \ldots & \ldots & \ldots & \ldots & 120 & 4 & 90 & 10 & none & Cha$^f$ \\
\object{HD 75393}  & F9 & 4.7 & 1.6 & a & \ldots & \ldots & \ldots & \ldots & 9 & 2 & $>$600 & 50 & none & field$^b$ \\
\object{RX J 0850.1-7554}  & G7 & 47.7 & 3.6 & a & \ldots & \ldots & \ldots & \ldots & 210 & 5 & 60 & 15 & none & Cha$^f$ \\
\object{RX J 0917.2-7744}  & G4.5 & 76.5 & 8.9 & a & \ldots & \ldots & \ldots & \ldots & 165 & 5 & 30 & 10 & none & Cha$^f$ \\
\object{HD 81544}$^\dagger$ & G9 & 12.9 & 0.5 & fi & \ldots & \ldots & \ldots & \ldots & 331 & 5 & 8 & 3 & \ldots & \ldots \\
\object{HD 84323}  & F9 & \ldots & \ldots & a & \ldots & \ldots & \ldots & \ldots & 197 & 5 & 30 & 15 & \ldots & field \\
\object{TYC 7697-2254-1}$^\dagger$ & G8 & 9.0 & 0.4 & fi & \ldots & \ldots & \ldots & \ldots & 240 & 4 & 25 & 10 & \ldots & \ldots \\
\object{HD 86356}  & G8 & 136.4 & 10.8 & -0.04 & 0.01 & 1.5 & \ldots & \ldots & 265 & 5 & 10 & 5 & none & Cha$^f$ \\
\object{HD 86021}$^\dagger$ & G7 & 24.3 & 1.6 & a & \ldots & \ldots & \ldots & \ldots & 78 & 10 & 200 & 100 & \ldots & \ldots \\
\object{TYC 6604-118-1}  & K1 & 18.3 & 1.1 & -0.11 & 0.01 & 1.5 & \ldots & \ldots & 72 & 5 & 8 & 5 & \ldots & TWA$^k$ \\
\object{TYC 6604-118-1B}  & \ldots & 65.4 & 6.0 & \ldots & \ldots & \ldots & \ldots & \ldots & \ldots & \ldots & \ldots & \ldots & \ldots & \ldots \\
\object{HD 88201}  & G0 & 10.3 & 0.9 & a & \ldots & \ldots & \ldots & \ldots & 71 & 4 & 300 & 50 & none & field$^b$ \\
\object{TWA 21}  & K2 & 8.2 & 1.0 & a & \ldots & \ldots & \ldots & \ldots & 357 & 3 & 8 & 3 & \ldots & TWA$^u$ \\
\object{TWA 6}  & K6 & 94.7 & 9.1 & -3.55 & 0.6 & 5.1 & \ldots & \ldots & 500 & 5 & 7 & 3 & non-accr.$^i$ & TWA$^u$ \\
\object{HD 90712}  & G1.5 & 11.2 & 0.9 & a & \ldots & \ldots & \ldots & \ldots & 99 & 1 & 250 & 50 & none & field$^b$ \\
\object{RX J 1029.5-6349}  & K1.5 & 24.3 & 1.5 & a & \ldots & \ldots & \ldots & \ldots & 251 & 5 & 40 & 15 & none & IC\,2602$^b$ \\
\object{RX J 1040.0-6315}  & G2.5 & 13.9 & 0.9 & a & \ldots & \ldots & \ldots & \ldots & 177 & 6 & 30 & 15 & none & IC\,2602$^b$ \\
\object{HD 307938}  & F9 & 81.1 & 5.0 & a & \ldots & \ldots & \ldots & \ldots & 161 & 1 & 30 & 15 & \ldots & IC\,2602$^b$ \\
\object{TWA 7}  & M1 & 5.0 & 2.1 & -4.64 & 0.5 & 2.1 & \ldots & \ldots & 465 & 10 & 8 & 3 & non-accr.$^i$ & TWA$^u$ \\
\object{RX J 1046.2-6402}  & G3 & 35.2 & 2.0 & a & \ldots & \ldots & \ldots & \ldots & 187 & 5 & 40 & 10 & none & IC\,2602$^b$ \\
\object{2MASS J10574936-6913599}  & K3 & 26.8 & 4.2 & a & \ldots & \ldots & \ldots & \ldots & 404 & 8 & 5 & 3 & none & ScoCen/LCC$^b$ \\
\object{V* CR Cha}$^\dagger$ & K2 & 36.5 & 4.4 & -26.94 & 1.24 & 10.55 & 0.06 & 0.07 & 380 & 3 & 5 & 3 & accr. & Cha \\
\object{V* TW Hya} & M0 & 6.2 & 1.3 & -116.78 & 5 & 9.5 & 1.94 & 0.8 & 361 & $<$1 & 8 & 3 & accr.$^e$ & TWA$^u$ \\
\object{HD 96064}  & G7 & 6.1 & 1.3 & a & \ldots & \ldots & \ldots & \ldots & 106 & 2 & 90 & 10 & \ldots & field$^t$ \\
\object{V* DI Cha}$^\dagger$ & G2 & 38.1 & 2.2 & -26.07 & 0.5 & 9.8 & 0.03 & 0.05 & 260 & 1 & 5 & 2 & accr.$^g$ & Cha \\
\object{TWA 2}  & M0.5 & 11.8 & 1.0 & -1.89 & 0.07 & 2.2 & \ldots & \ldots & 509 & 11 & 8 & 3 & non-accr.$^i$ & TWA$^u$ \\
\object{HD 98553}  & G2 & 4.0 & 1.9 & a & \ldots & \ldots & \ldots & \ldots & 21 & 1 & 300 & 50 & none & field$^b$ \\
\object{TWA 4}  & K1 & 5.0 & 1.6 & -0.09 & 0.01 & 1.5 & \ldots & \ldots & 312 & 5 & 8 & 5 & non-accr.$^i$ & TWA$^u$ \\
\object{TWA 4B}  & \ldots & 8.0 & 1.1 & \ldots & \ldots & \ldots & \ldots & \ldots & \ldots & \ldots & \ldots & \ldots & \ldots & \ldots \\
\object{HD 99409}  & G6 & 17.6 & 1.1 & a & \ldots & \ldots & \ldots & \ldots & 95 & $<$1 & 125 & 20 & \ldots & field$^t$ \\
\object{TWA 5}  & M1.5 & 63.2 & 9.5 & -8.5 & 0.3 & 8.8 & \ldots & \ldots & 564 & 5 & 8 & 3 & none$^i$ & TWA$^u$ \\
\object{TWA 8}  & M1 & 7.1 & 1.2 & -6.43 & 0.5 & 2.8 & \ldots & \ldots & 475 & 5 & 8 & 3 & non-accr.$^i$ & TWA$^u$ \\
\object{RX J 1140.3-8321}  & K1 & 9.3 & 1.1 & -0.18 & 0.02 & 2.1 & \ldots & \ldots & 195 & 1 & 15 & 10 & none & Cha$^f$ \\
\object{HD 101472}  & F8.5 & 5.8 & 1.3 & a & \ldots & \ldots & \ldots & \ldots & 112 & 2 & 90 & 10 & none & field$^b$ \\
\object{TWA 19}  & G7 & 30.1 & 1.7 & a & \ldots & \ldots & \ldots & \ldots & 197 & 5 & 8 & 5 & non-accr.$^i$ & TWA$^u$ \\
\object{V* GQ Leo} & K8 & 7.2 & 1.6 & -0.63 & 0.02 & 2.2 & \ldots & \ldots & 83 & 2 & 12 & 4 & none & field \\
\object{TWA 9}  & K4 & 10.4 & 1.0 & -1.57 & 0.07 & 2.1 & \ldots & \ldots & 506 & 5 & 7 & 3 & non-accr.$^i$ & TWA$^u$ \\
\object{V* T Cha}$^\dagger$ & G4 & 40.0 & 2.7 & fi & \ldots & \ldots & \ldots & \ldots & 382 & 1 & 4 & 2 & non-accr.$^a$ & Cha \\
\object{HD 104576}  & G1 & 3.5 & 2.1 & a & \ldots & \ldots & \ldots & \ldots & 52 & 2 & 300 & 50 & none & field$^b$ \\
\object{RX J 1203.7-8129}  & K2 & 2.7 & 2.7 & a & \ldots & \ldots & \ldots & \ldots & 69 & 6 & 250 & 50 & none & Cha$^f$? \\
\object{TWA 24}  & K1.5 & 15.5 & 1.0 & -0.6 & 0.02 & 2.8 & \ldots & \ldots & 383 & 5 & 7 & 3 & \ldots & TWA$^u$ \\
\object{HD 105690}  & G4 & 8.5 & 1.0 & a & \ldots & \ldots & \ldots & \ldots & 149 & 1 & 8 & 15 & \ldots & TWA$^k$ \\
\object{2MASS J12123577-5520273}  & K4 & 28.5 & 1.6 & -0.2 & 0.02 & 1.5 & \ldots & \ldots & 358 & 2 & 5 & 3 & debris & ScoCen/LCC$^b$ \\
\object{2MASS J12143410-5110124}  & K0 & 17.9 & 1.1 & a & \ldots & \ldots & \ldots & \ldots & 311 & 5 & 25 & 10 & \ldots & ScoCen/LCC$^b$ \\
\object{TWA 25}  & M0 & 11.6 & 1.1 & -1.87 & 0.6 & 2.8 & \ldots & \ldots & 531 & 13 & 8 & 3 & \ldots & TWA$^u$ \\
\object{RX J 1220.6-7539}  & K2.5 & 6.3 & 1.3 & -0.2 & 0.03 & 2.1 & \ldots & \ldots & 208 & 2 & 15 & 10 & none & Cha$^f$ \\
\object{HD 107441}  & K2 & 41.5 & 2.7 & a & \ldots & \ldots & \ldots & \ldots & 226 & 4 & 30 & 10 & none & ScoCen/LCC$^b$ \\
\object{2MASS J12223322-5333489}  & G1 & 70.9 & 7.3 & a & \ldots & \ldots & \ldots & \ldots & 191 & 5 & 30 & 10 & debris & ScoCen/LCC$^b$ \\
\object{RX J 1225.3-7857}   & G7 & 8.2 & 1.0 & a & \ldots & \ldots & \ldots & \ldots & 119 & 5 & 125 & 20 & none & Cha$^f$ \\
\object{HD 108799}  & G0 & 7.2 & 1.1 & a & \ldots & \ldots & \ldots & \ldots & 94 & 2 & 125 & 20 & none & field$^b$ \\
\object{HD 109138}$^\dagger$ & K0 & 6.4 & 0.4 & a & \ldots & \ldots & \ldots & \ldots & 129 & 3 & 90 & 15 & \ldots & \ldots \\
\object{1RXS J123332.4-571345}$^\dagger$ & K1 & 18.2 & 0.8 & fi & \ldots & \ldots & \ldots & \ldots & 424 & 8 & 8 & 5 & \ldots & \ldots \\
\object{TYC 9412-59-1}$^\dagger$ & K3 & 18.0 & 0.8 & fi & \ldots & \ldots & \ldots & \ldots & 429 & 8 & 5 & 3 & \ldots & Cha \\
\object{TYC 8654-1115-1}$^\dagger$ & G9 & 19.4 & 0.4 & a & \ldots & \ldots & \ldots & \ldots & 298 & 7 & 5 & 3 & \ldots & \ldots \\
\object{HD 111170}  & G9 & 133.2 & 7.7 & a & \ldots & \ldots & \ldots & \ldots & 4 & 1 & $>$600 & 50 & none & ScoCen/LCC$^b$ \\
\object{TYC 8655-149-1}  & G7.5 & 28.0 & 1.6 & a & \ldots & \ldots & \ldots & \ldots & 235 & 5 & 30 & 10 & none & ScoCen/LCC$^b$ \\
\object{CD-69 1055}$^\dagger$ & K0 & 26.0 & 1.5 & fi & \ldots & \ldots & \ldots & \ldots & 400 & 6 & 5 & 3 & \ldots & LCC \\
\object{2MASS J13015069-5304581}  & K1.5 & 8.9 & 1.0 & a & \ldots & \ldots & \ldots & \ldots & 367 & 5 & 8 & 3 & debris & ScoCen/LCC$^b$ \\
\object{HD 113553}  & G4 & 11.6 & 0.9 & a & \ldots & \ldots & \ldots & \ldots & 139 & 1 & 90 & 10 & \ldots & field$^r$ \\
\object{TYC 8649-251-1}  & F9.5 & 115.8 & 10.2 & a & \ldots & \ldots & \ldots & \ldots & 208 & 5 & 30 & 15 & none & ScoCen/LCC$^b$ \\
\object{V* MP Mus} & K2 & 12.9 & 1.0 & -42.11 & 3 & 10.5 & 0.5 & 0.1 & 410 & 5 & 7 & 3 & accr. & ScoCen/LCC$^b$ \\
\object{HD 117524}  & G7 & 32.1 & 1.9 & a & \ldots & \ldots & \ldots & \ldots & 210 & 3 & 30 & 10 & none & ScoCen/LCC$^b$ \\
\object{2MASS J13375730-4134419}  & K1 & 13.7 & 1.0 & a & \ldots & \ldots & \ldots & \ldots & 312 & 3 & 30 & 10 & debris & ScoCen/UCL$^b$ \\
\object{HD 119269}  & G5 & 24.2 & 1.4 & a & \ldots & \ldots & \ldots & \ldots & 211 & 4 & 30 & 10 & \ldots & ScoCen/LCC$^b$ \\
\object{TYC 8270-2015-1}  & K0 & 69.6 & 8.7 & -0.17 & 0.02 & 1.8 & \ldots & \ldots & 232 & 5 & 30 & 10 & \ldots & ScoCen/UCL$^b$ \\
\object{HD 120812}  & F9.5 & 25.9 & 1.5 & a & \ldots & \ldots & \ldots & \ldots & 94 & 5 & 30 & 15 & none & ScoCen/UCL$^b$ \\
\object{TYC 7811-2909-1}  & K1 & 17.5 & 1.1 & a & \ldots & \ldots & \ldots & \ldots & 329 & 5 & 25 & 10 & none & ScoCen$^b$ \\
\object{HD 124784}$^\dagger$ & G0 & 17.3 & 1.1 & a & \ldots & \ldots & \ldots & \ldots & 140 & 1 & 200 & 50 & \ldots & UCL \\
\object{HD 124784B}  & \ldots & 17.5 & 1.1 & \ldots & \ldots & \ldots & \ldots & \ldots & \ldots & \ldots & \ldots & \ldots & \ldots & \ldots \\
\object{TYC 8282-516-1}  & G9 & 10.3 & 0.9 & a & \ldots & \ldots & \ldots & \ldots & 290 & 5 & 30 & 10 & debris & ScoCen/UCL$^b$ \\
\object{HD 126670}  & G7 & 13.4 & 0.9 & a & \ldots & \ldots & \ldots & \ldots & 275 & 2 & 30 & 10 & none & ScoCen/UCL$^b$ \\
\object{HD 128242}  & G1 & 115.8 & 6.7 & a & \ldots & \ldots & \ldots & \ldots & 60 & 12 & 300 & 50 & none & ScoCen/UCL$^b$ \\
\object{RX J 1457.3-3613}  & G5 & 31.9 & 1.8 & a & \ldots & \ldots & \ldots & \ldots & 280 & 3 & 20 & 10 & none & ScoCen/UCL$^b$ \\
\object{HD 132173}  & F9 & 9.6 & 0.9 & a & \ldots & \ldots & \ldots & \ldots & 117 & 2 & 30 & 10 & \ldots & field$^b$ \\
\object{TYC 8297-1613-1}  & G4 & 17.9 & 1.1 & a & \ldots & \ldots & \ldots & \ldots & 192 & 5 & 30 & 10 & \ldots & ScoCen/UCL$^b$ \\
\object{RX J 1507.2-3505}   & K0.5 & 12.2 & 0.9 & a & \ldots & \ldots & \ldots & \ldots & 320 & 5 & 30 & 10 & none & ScoCen/UCL$^b$ \\
\object{HD 133938}  & G7.5 & 24.9 & 1.4 & a & \ldots & \ldots & \ldots & \ldots & 223 & 4 & 45 & 15 & none & ScoCen/UCL$^b$ \\
\object{V* LT Lup}$^\dagger$ & K0 & 27.8 & 1.7 & fi & \ldots & \ldots & \ldots & \ldots & 385 & 5 & 8 & 3 & none$^o$ & Lup \\
\object{RX J 1518.4-3738}   & K2.5 & 25.2 & 1.5 & a & \ldots & \ldots & \ldots & \ldots & 321 & 5 & 30 & 15 & none & ScoCen/UCL$^b$ \\
\object{V* LY Lup}$^\dagger$ & K0 & 17.2 & 1.1 & -0.5 & 0.05 & 2.5 & \ldots & \ldots & 401 & 10 & 3 & 2 & none$^o$ & Lup \\
\object{V* MP Lup}$^\dagger$ & K0 & 18.0 & 1.3 & fi & \ldots & \ldots & \ldots & \ldots & 360 & 2 & 5 & 3 & none$^o$ & Lup \\
\object{V* MS Lup}$^\dagger$ & G7 & 28.9 & 1.8 & a & \ldots & \ldots & \ldots & \ldots & 308 & 6 & 5 & 3 & none$^o$ & Lup \\
\object{RX J 1531.3-3329}   & G9 & 15.6 & 1.0 & a & \ldots & \ldots & \ldots & \ldots & 281 & 5 & 30 & 15 & none & ScoCen/UCL$^l$ \\
\object{1RXS J153328.4-665130}\footnote{{\scriptsize \object{TYC 9034-968-1}}}$^\dagger$ & K2 & 9.1 & 1.1 & fi & \ldots & \ldots & \ldots & \ldots & 295 & 1 & 30 & 15 & \ldots & field \\
\object{HD 139084}  & G8 & 18.2 & 1.1 & a & \ldots & \ldots & \ldots & \ldots & 278 & 4 & 50 & 15 & none & $\beta$\,Pic$^u$ \\
\object{HD 139498}  & G7 & 32.6 & 3.2 & a & \ldots & \ldots & \ldots & \ldots & 149 & 5 & 30 & 10 & none & ScoCen/UCL$^b$ \\
\object{HD 139498B}  & \ldots & 17.4 & 1.1 &  & \ldots & \ldots & \ldots & \ldots & \ldots & \ldots & \ldots & \ldots & \ldots & \ldots \\
\object{HD 140374}  & G7.5 & 5.1 & 1.5 & a & \ldots & \ldots & \ldots & \ldots & 124 & 5 & 30 & 10 & none & ScoCen/UCL$^b$ \\
\object{HD 140374B}  & \ldots & 8.1 & 1.1 &  & \ldots & \ldots & \ldots & \ldots & \ldots & \ldots & \ldots & \ldots & \ldots & \ldots \\
\object{RX J 1544.0-3311}  & K1 & 10.3 & 1.0 & -1.25 & 0.03 & 3.7 & \ldots & \ldots & 357 & 3 & 12 & 3 & none & ScoCen/UCL$^b$ \\
\object{V* HT Lup}$^\dagger$ & K2 & 44.8 & 3.0 & -7.02 & 0.2 & 11.0 & \ldots & \ldots & 442 & 5 & 5 & 2 & non-accr. & Lup \\
\object{V* GQ Lup}$^\dagger$ & K8 & 9.1 & 1.1 & -39.53 & 0.5 & 13.2 & 0.6 & 0.5 & 644 & 5 & 2 & 3 & accr. & Lup \\
\object{HD 141521} & G3.5 & 7.1 & 1.1 & a & \ldots & \ldots & \ldots & \ldots & 149 & 2 & 30 & 10 & none & ScoCen/UCL$^b$ \\
\object{HD 141521B}  & \ldots & 8.1 & 1.1 & \ldots & \ldots & \ldots & \ldots & \ldots & \ldots & \ldots & \ldots & \ldots & \ldots & \ldots \\
\object{HBC 603}$^\dagger$ & M0 & 5.9 & 0.6 & -16.08 & 2.5 & 7.71 & 0.1 & 0.05 & 566 & 10 & 5 & 3 & accr. & \ldots \\
\object{HD 141943}  & G4 & 36.3 & 2.1 & a & \ldots & \ldots & \ldots & \ldots & 198 & 2 & 30 & 15 & debris & field$^b$ \\
\object{HD 142229}  & G0 & 5.6 & 1.4 & a & \ldots & \ldots & \ldots & \ldots & 116 & 5 & 90 & 10 & none & field$^b$ \\
\object{HD 142361}  & G3 & 79.4 & 4.6 & a & \ldots & \ldots & \ldots & \ldots & 195 & 5 & 30 & 10 & none & ScoCen/UCL$^b$ \\
\object{V* IM Lup}$^\dagger$ & M0 & 12.6 & 0.6 & -5.71 & 0.51 & 5.5 & 0.03 & 0.1 & 598 & 10 & 5 & 3 & accr. & Lup \\
\object{HD 143358}  & G1 & 72.5 & 4.4 & a & \ldots & \ldots & \ldots & \ldots & 166 & 3 & 30 & 10 & \ldots & ScoCen/UCL$^b$ \\
\object{EM* SR 9}$^\dagger$ & K5 & 13.4 & 0.7 & -20.38 & 1.42 & 7.82 & 0.3 & 0.15 & 567 & 8 & 5 & 3 & accr. & \ldots \\
\object{V* V2129 Oph}$^\dagger$ & K3 & 13.5 & 1.0 & -14.6 & 2 & 6.5 & 0.25 & 0.15 & 530 & 5 & 3 & 2 & accr. & Oph \\
\object{DoAr 44}$^\dagger$ & K3 & 14.6 & 1.1 & -22.87 & 4.42 & 9.98 & 0.6 & 0.15 & 507 & 8 & 5 & 3 & accr. & \ldots \\
\object{V* V1121 Oph}$^\dagger$ & K5 & 8.4 & 0.5 & -20.81 & 2.48 & 8.28 & 0.3 & 0.25 & 550 & 6 & 8 & 3 & accr. & Oph \\
\object{HD 151798}  & G1 & 11.3 & 0.9 & a & \ldots & \ldots & \ldots & \ldots & 131 & 2 & 90 & 10 & none & field$^b$ \\
\object{HD 152555}  & F9 & 17.8 & 1.1 & a & \ldots & \ldots & \ldots & \ldots & 123 & 2 & 50 & 10 & none & AB\,Dor$^u$ \\
\object{TYC 8726-57-1}$^\dagger$ & G9 & 22.1 & 1.4 & a & \ldots & \ldots & \ldots & \ldots & 250 & 4 & 25 & 15 & \ldots & field \\
\object{CD-27 11535}$^\dagger$ & K5 & 12.3 & 1.1 & -1.06 & 0.07 & 3.1 & \ldots & \ldots & 490 & 1 & 6 & 3 & \ldots & field \\
\object{TYC 6812-348-1}$^\dagger$ & K3 & 19.1 & 1.4 & -0.27 & 0.05 & 2.0 & \ldots & \ldots & 455 & 4 & 5 & 2 & \ldots & field \\
\object{HD 155555}  & K2 & 39.9 & 2.3 & a & \ldots & \ldots & \ldots & \ldots & 170 & 5 & 12 & 3 & none$^p$ & $\beta$\,Pic$^u$ \\
\object{HD 155555B}  & \ldots & 8.8 & 0.9 & \ldots & \ldots & \ldots & \ldots & \ldots & 169 & 5 & \ldots & \ldots & \ldots & \ldots \\
\object{HD 161460}$^\dagger$ & K0 & 10.1 & 1.0 & fi & \ldots & \ldots & \ldots & \ldots & 320 & 4 & 30 & 15 & \ldots & field \\
\object{RX J 1841.8-3525}  & G7 & 41.2 & 2.3 & a & \ldots & \ldots & \ldots & \ldots & 248 & 1 & 30 & 15 & none & Cor\,Aus$^b$  \\
\object{HD 174656}  & G4 & 27.8 & 1.9 & a & \ldots & \ldots & \ldots & \ldots & 338 & 5 & 8 & 3 & none & Cor\,Aus$^b$ \\
\object{HD 181327}  & F5.5 & 19.4 & 1.2 & a & \ldots & \ldots & \ldots & \ldots & 108 & $<$1 & 15 & 10 & debris$^p$ & $\beta$\,Pic$^u$ \\
\object{HD 183414}  & G3 & 9.6 & 0.9 & a & \ldots & \ldots & \ldots & \ldots & 142 & 3 & 75 & 15 & \ldots & field$^r$ \\
\object{HD 191089}  & F9 & 48.8 & 3.0 & a & \ldots & \ldots & \ldots & \ldots & 101 & 5 & 125 & 20 & \ldots & field$^b$ \\
\object{HD 193464}  & F9 & 15.0 & 1.0 & a & \ldots & \ldots & \ldots & \ldots & 92 & 2 & 300 & 50 & \ldots & field$^t$ \\
\object{HD 197481}  & M0.5 & 8.2 & 1.2 & -3.16 & 0.05 & 2.8 & \ldots & \ldots & 56 & 3 & 15 & 10 & debris$^p$ & $\beta$\,Pic$^u$ \\
\object{HD 199143}  & F9.5 & 161.5 & 11.8 & a & \ldots & \ldots & \ldots & \ldots & 60 & 5 & 300 & 50 & none & $\beta$\,Pic$^u$ \\
\object{HD 200746}  & G1.5 & 6.4 & 1.2 & a & \ldots & \ldots & \ldots & \ldots & 78 & 5 & 300 & 50 & none & field$^b$ \\
\object{HD 201219}  & G2.5 & 3.5 & 2.1 & a & \ldots & \ldots & \ldots & \ldots & n/a & \ldots & \ldots & \ldots & debris & field$^b$ \\
\object{HD 201989}  & G2 & 5.0 & 1.5 & \ldots & \ldots & \ldots & \ldots & \ldots & n/a & \ldots & \ldots & \ldots & none & field$^b$ \\
\object{HD 202917}  & G4.5 & 14.9 & 1.0 & a & \ldots & \ldots & \ldots & \ldots & 211 & 1 & 30 & 15 & \ldots & Tuc/Hor$^u$ \\
\object{HD 202947}  & K1 & 12.9 & 0.9 & a & \ldots & \ldots & \ldots & \ldots & 161 & 11 & 30 & 10 & debris$^m$ & Tuc/Hor$^u$ \\
\object{HD 205905}  & K1 & 4.8 & 1.6 & \ldots & \ldots & \ldots & \ldots & \ldots & n/a & \ldots & \ldots & \ldots & none & field$^b$ \\
\object{HD 207278}$^\dagger$ & G7 & 11.2 & 0.9 & a & \ldots & \ldots & \ldots & \ldots & 190 & 1 & 50 & 10 & \ldots & AB\,Dor \\
\object{HD 207575}  & F9 & 36.5 & 2.1 & a & \ldots & \ldots & \ldots & \ldots & 111 & 7 & 30 & 10 & debris$^m$ & Tuc/Hor$^u$ \\
\object{HD 209253}  & F8 & 16.9 & 1.0 & \ldots & \ldots & \ldots & \ldots & \ldots & n/a & \ldots & \ldots & \ldots & \ldots & field$^b$ \\
\object{HD 209779}  & G0 & 5.5 & 1.4 & a & \ldots & \ldots & \ldots & \ldots & n/a & \ldots & \ldots & \ldots & none & field$^b$ \\
\object{HD 212291}  & G4 & 1.5 & 4.4 & a & \ldots & \ldots & \ldots & \ldots & n/a & \ldots & \ldots & \ldots & none & field$^b$ \\
\object{1RXS J223929.1-520525}$^\dagger$ & K0 & 9.4 & 1.0 & a & \ldots & \ldots & \ldots & \ldots & 245 & 1 & 60 & 15 & \ldots & field \\
\object{CP-72 2713}$^\dagger$ & K7 & 6.6 & 1.4 & -12.5 & 0.6 & 2.3 & \ldots & \ldots & 440 & 4 & 8 & 1 & \ldots & field \\
\object{CD-40 14901}$^\dagger$ & G5 & 14.9 & 1.0 & a & \ldots & \ldots & \ldots & \ldots & 175 & 2 & 75 & 15 & \ldots & field \\
\object{HD 216803}  & K3 & 4.5 & 1.8 & a & \ldots & \ldots & \ldots & \ldots & 26 & 5 & 400 & 50 & \ldots & field$^b$ \\
\object{HD 217343}  & G3 & 12.2 & 0.9 & a & \ldots & \ldots & \ldots & \ldots & 163 & 1 & 45 & 10 & none & AB\,Dor$^u$ \\
\object{HD 217897}$^\dagger$ & M2.5 & 8.5 & 1.2 & a & \ldots & \ldots & \ldots & \ldots & 50 & 3 & 30 & 15 & \ldots & field \\
\object{BD-03 5579}  & K4 & 14.0 & 1.0 & -0.7 & 0.02 & 1.3 & \ldots & \ldots & 200 & 3 & 50 & 15 & \ldots & field \\
\object{TYC 9129-1361-1}  & K1 & 86.1 & 9.8 & a & \ldots & \ldots & \ldots & \ldots & $<$2 & 5 & $>$600 & 50 & \ldots & field \\
\object{TYC 9129-1361-1B}  & \ldots & 119.5 & 10.0 & \ldots & \ldots & \ldots & \ldots & \ldots & \ldots & \ldots & \ldots & \ldots & \ldots & \ldots \\
\object{HD 220054}  & G8 & 31.8 & 1.9 & a & \ldots & \ldots & \ldots & \ldots & 259 & 4 & 30 & 10 & \ldots & field \\
\object{TYC 584-343-1}$^\dagger$ & K0 & 14.0 & 1.0 & fi & \ldots & \ldots & \ldots & \ldots & 300 & 2 & 30 & 15 & \ldots & field \\
\object{BD-13 6424}$^\dagger$ & M0 & 9.6 & 1.2 & -1.2 & 0.03 & 1.9 & \ldots & \ldots & 185 & 8 & 30 & 15 & \ldots & field \\
\object{HD 222259}  & G4 & 17.5 & 1.1 & a & \ldots & \ldots & \ldots & \ldots & 204 & 5 & 30 & 10 & debris$^m$ & Tuc/Hor$^u$ \\
\object{HD 224228}  & K2 & 2.7 & 2.8 & a & \ldots & \ldots & \ldots & \ldots & 60 & 5 & 50 & 20 & \ldots & AB\,Dor$^u$ \\

\hline
\end{longtable}
\end{landscape}
}}
%\end{longtabL}
%\appendix
\begin{appendix}
\section{Calibration of $v\sin i$}
\label{sec_vm}

In this Appendix, the calibration for the $v\sin i$ measurements in Section~\ref{sec_vm1} is presented. The $v\sin i$ can be calculated from the width of the CCF $\sigma_\mathrm{CCF}$ by using
\begin{equation}
\label{eq_vsini}
v\sin{i} = A \sqrt{\sigma_{\mathrm{CCF}}^{2}-\sigma_{0}^2} \, ,
\end{equation}
where $A$ is a coupling constant and $\sigma_0$ is a quantity for intrinsic stellar line-broadening effects (see Section~\ref{sec_vm1}).

The coupling constant $A$ and $\sigma_0$ can be calibrated independently of each other, since the quantity $A$ only depends on instrumental and numerical parameters and the quantity $\sigma_0$ depends on intrinsic stellar effects. On the other hand, each is needed to calibrate the other variable. To be able to calibrate $A$, $\sigma_0$ has been taken as the width of the unbroadenend artifical spectra (see Section~\ref{sec_cala}). The resulting $A$ has been used to calibrate $\sigma_0$ (Section~\ref{sec_calibration}) with spectra of slow rotating stars with different effective temperatures. As a cross-check, $\sigma_0$ computed in Section~\ref{sec_calibration} has been iteratively used to recalculate $A$ (Section~\ref{sec_cala}), which yielded the same results for $A$.

\subsection{Calibration of the coupling constant A}
\label{sec_cala}
The coupling constant A has been derived in a similar way as described by Queloz et al. (1998). For this analysis, theoretical spectra have been cross-correlated with an appropriate template. A synthetic stellar spectrum with effective temperature of 5750\,K and $\log g = 4.5$\,dex has been produced from Kurucz models (Kurucz 1993), using the program SPECTRUM (Gray \& Corbally 1994). This has been done to have the best agreement with the used template for G-type stars. In a first step, the theoretical spectrum has been convolved with a Gaussian-shape instrumental profile matching that of FEROS or HARPS, respectively. Then, the theoretical spectrum with the instrumental profile has been broadened to $v\sin i$ = 0, 5, 10, 15, 20, 25 and 30 kms$^{-1}$ and the respective CCF has been calculated by using MACS. The widths of the CCF obtained by a Gaussian fit, $\sigma_{\mathrm{CCF}}$, are listed in Table~\ref{tab_calA}. The effective line width of a non-rotating star, $\sigma_{0}$, has been adopted to be equal to the width of the CCF with $v\sin i = 0\,\mathrm{kms}^{-1}$.

The coupling constant $A$ was then determined as the weighted mean of the $A$-values derived for the CCFs with different $v\sin i$, using Equation~\ref{eq_vsini}.
We derive
\begin{equation}
A_{\mathrm{FEROS}} = 1.8\,\pm\,0.1
\end{equation}
and
\begin{equation}
A_{\mathrm{HARPS}} = 1.88 \pm 0.05 ~.
\end{equation}

In both cases, the cross-check by using the $\sigma_0$ from Section~\ref{sec_calibration} yielded the same result for $A$.

%also applicable to stars with $v\sin i \leq$\,30\,kms$^{-1}$, as well.
\begin{table}[th]
\caption{Calibration of A}
$$
\vspace{-1.1cm}
$$
$$
\begin{array}{cccc}
\hline
\hline
\\
v\sin i & \sigma_{\mathrm{CCF}} & \sqrt{\sigma_{\mathrm{CCF}}^2-\sigma_{0}^2} & A\\
\mathrm{kms}^{-1} & \mathrm{kms}^{-1} & \mathrm{kms}^{-1} & \\
\hline
\\
\multicolumn{4}{c}{FEROS}\\
%\\
\hline
\\
0 & 4.47 \pm 0.01 & \ldots & \ldots \\
5 & 5.47 \pm 0.02 & 3.15 \pm 0.02 & 1.59 \pm 0.02\\
10 & 6.85 \pm 0.04 & 5.15 \pm 0.04 & 1.96 \pm 0.03\\
15 & 9.69 \pm 0.07 & 8.60 \pm 0.08 & 1.75 \pm 0.05\\
20 & 11.07 \pm 0.10 & 10.13 \pm 0.09 & 1.98 \pm 0.03\\
25 & 14.01 \pm 0.25 & 13.28 \pm 0.27 & 1.88 \pm 0.16\\
30 & 17.34 \pm 0.44 & 16.75 \pm 0.48 & 1.79 \pm 0.30\\
\hline
\\
\multicolumn{4}{c}{HARPS}\\
%\\
\hline
\\
%v\sin i & \sigma_{\mathrm{CCF}} & \sqrt{\sigma_{\mathrm{CCF}}^2-\sigma_{0}^2} & A\\
%\mathrm{km\,s}^{-1} & \mathrm{km\,s}^{-1} & \mathrm{km\,s}^{-1} & \\
%\hline
%\\
0 & 3.86 \pm 0.01 & \ldots & \ldots \\
5 & 4.56 \pm 0.02 & 2.45 \pm 0.02  & 2.05 \pm 0.02\\
10 & 6.54 \pm 0.03 & 5.28 \pm 0.02 & 1.89 \pm 0.01\\
15 & 9.02 \pm 0.07 & 8.15 \pm 0.06 & 1.84 \pm 0.05\\
20 & 11.81 \pm 0.13 & 11.16 \pm 0.12 & 1.79 \pm 0.09\\
25 & 13.85 \pm 0.23 & 13.30 \pm 0.22 & 1.88 \pm 0.15\\
30 & 16.62 \pm 0.39 & 16.17 \pm 0.37 & 1.86 \pm 0.33\\
\hline
\hline
\end{array}
$$
\label{tab_calA}
\end{table}

The scatter of $A$ around the adopted weighted mean value is due to the fact that the shape of rotationally broadened spectral lines is no longer well represented by a Gaussian (e.g., Gray 1992). Thus, the scatter has been introduced by the imperfect matching of the Gaussian fitting function. This behaviour only affects the measurement of the width of the CCF but not the determination of the minimum position, since the CCF of the theoretical spectra is symmetric.
In order to test, whether other fitting functions describe the resulting CCF better, several line-profile functions have been used, like an additionally broadened Gaussian. This function has been adopted from Hirano et al. (2010). They describe the stellar rotation profile by a Gaussian, which is then convolved with the first Gaussian used to describe the unbroadened spectral line.\\
As a result, measurements with all fitting functions yielded the same mean value for $A$ and the scatter around it is not significantly reduced compared to a single Gaussian fit. Thus, an ideal fitting function for a rotationally broadened CCF is not yet existent.

\subsection{Calibration of $\sigma_0$}
\label{sec_calibration}
\begin{figure}
\centering
\includegraphics[width=8.0cm]{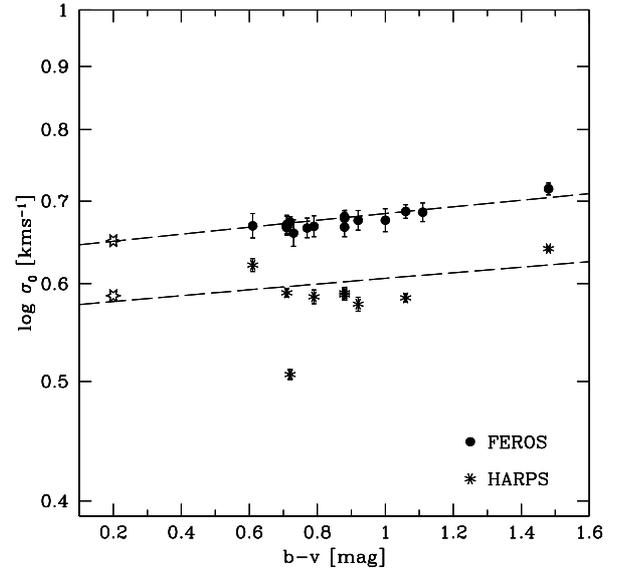}
\caption{Calibration of $\sigma_{0}$ using slowly rotating stars listed in Table~\ref{tab_vsinikalib}. Also shown at $(\mathrm{b}-\mathrm{v}) = 0.2$ (open stars) is the $\sigma_0$ measured in Section~\ref{sec_cala}. The dashed lines show the fit to the data.}
\end{figure}
In order to correctly measure the projected rotational velocity $v\sin i$ of a target and to disentangle it from the intrinsic and instrumental broadening effects, $\sigma_{0}$ has to be known. The instrumental broadening effects should be independent of stellar properties of the observed star, while the intrinsic stellar line broadening depends strongly on $T_\mathrm{eff}$, due to Doppler broadening. From the Maxwellian most likely velocity of atoms in thermal motion, $\mathrm{v}^2_0 \propto T$, and from the relation of the Doppler wavelength shift to this velocity, $\Delta \lambda = \mathrm{v}_{0}\lambda/c$, it follows that
\begin{equation}
\label{eq_cal1}
\Delta{}\lambda \propto \sqrt{T_\mathrm{eff}} .
\end{equation}
Here $\Delta{}\lambda$ is equivalent to $\sigma_0 = \sigma_{0,\star}$.

Since not all stars have reliable measurements of $T_\mathrm{eff}$, but colours, like $(\mathrm{b}-\mathrm{v})$ are available and can directly be linked to $T_\mathrm{eff}$, the relation of line broadening with $(\mathrm{b}-\mathrm{v})$ has been investigated. According to Flower (1996), the relation between $(\mathrm{b}-\mathrm{v})$ and $\log T$ for main-sequence stars and giants is linear within the region of interest of $0<(\mathrm{b}-\mathrm{v})<1.6$, such that
\begin{equation}
\label{eq_cal2}
(\mathrm{b}-\mathrm{v}) \propto \log T_\mathrm{eff} ~[\mathrm{K}]
\end{equation}
holds.

Hence, we can use
\begin{equation}
\label{eq_sigma0}
\log \sigma_0 = a_0 + a_1 \cdot (\mathrm{b}-\mathrm{v})% \quad \mathrm{kms}^{-1} \quad .
%\sigma_{0} = a_0 + a_1 \cdot(\mathrm{b}-\mathrm{v})+ a_2 \cdot(\mathrm{b}-\mathrm{v})^2 \quad \mathrm{kms}^{-1} \,.
\end{equation}
to fit the data. In order to determine this relation, slowly rotating stars have been observed with FEROS at the 2.2m MPG/ESO telescope in La Silla, Chile in 2003 and 2004 and with HARPS at the 3.6m telescope in La Silla, Chile in 2008 and 2009, see Table~\ref{tab_vsinikalib}.
\begin{table}[h]
\caption{Calibration stars for $\sigma_{0}$, sorted by ($\mathrm{b}-\mathrm{v}$)}
$$
\vspace{-1.1cm}
$$
$$
\begin{array}{lcccll}
\hline
\hline
\\
\mathrm{Identifier} & (\mathrm{b}-\mathrm{v}) & v\sin{i}_{\mathrm{Valenti}}  & \mathrm{Instrument}   & \sigma_{CCF}& \sigma_{0} \\
\mathrm{}           & \mathrm{mag}    & \mathrm{km\,s}^{-1} & &  \mathrm{km\,s}^{-1}  & \mathrm{km\,s}^{-1}  \\
\hline
\\
\mathrm{HD}\, 76932 &	0.53 & 	2.6^a & \mathrm{FEROS} & 4.97 \pm 2.77 &	4.76 \pm 2.89\\
\mathrm{HD}\, 1581  &	0.61 &	3.0 & \mathrm{FEROS} & 4.95 \pm 0.04 &	4.66 \pm 0.11\\
\mathrm{HD}\, 1581 & 0.61 &	3.0 & \mathrm{HARPS} & 4.47\pm0.04 & 4.18 \pm 0.05 \\
\mathrm{HD}\, 20794 &	0.71 &	1.5 & \mathrm{FEROS} &	4.71 \pm 0.03 &	4.63 \pm 0.06\\
\mathrm{HD}\, 20794 & 0.71 &	1.5 & \mathrm{HARPS} & 3.97\pm0.03 & 3.89 \pm 0.03\\
\mathrm{HD}\,115617 &	0.71 &	2.2 & \mathrm{FEROS} &	4.82 \pm 0.04 &	4.66 \pm 0.08\\
\mathrm{HD}\,128620 &	0.71 &	2.3 & \mathrm{FEROS} &	4.85 \pm 0.03 &	4.68 \pm 0.08\\
\mathrm{HD}\, 10700 &	0.72 &	1.3 & \mathrm{FEROS} &	4.77 \pm 0.02 &	4.72 \pm 0.05\\
\mathrm{HD}\, 10700 & 0.72 &	1.3 & \mathrm{HARPS} & 3.28\pm0.03 & 3.21 \pm 0.03\\
\mathrm{HD}\, 73752 &	0.73 &	3.3^a & \mathrm{FEROS} &	4.92 \pm 0.02 &	4.63 \pm 0.11\\
\mathrm{HD}\, 13445 &	0.77 &	2.4 & \mathrm{FEROS} &	4.82 \pm 0.03 &	4.63 \pm 0.09\\
\mathrm{HD}\, 72673 &	0.79 &	2.7^a & \mathrm{FEROS} &	4.89 \pm 0.02 &	4.65 \pm 0.09\\
\mathrm{HD}\, 72673 & 0.79 &	2.7^a & \mathrm{HARPS} & 4.10\pm0.03 & 3.85 \pm 0.05\\
\mathrm{HD}\, 4628  &	0.88 &	2.0 & \mathrm{FEROS} &	4.89 \pm 0.03 &	4.76 \pm 0.07\\
\mathrm{HD}\, 4628 & 0.88 &	2.0 & \mathrm{HARPS} & 4.01\pm0.02 & 3.87 \pm 0.04\\
\mathrm{HD}\, 22049 &	0.88 &	2.4 & \mathrm{FEROS} &	4.83 \pm 0.02 &	4.64 \pm 0.09\\
\mathrm{HD}\, 22049 & 0.88 &	2.4 & \mathrm{HARPS} & 4.09\pm0.02 & 3.89 \pm 0.04\\
\mathrm{HD}\,128621 &	0.88 &	0.9 & \mathrm{FEROS} &	4.81 \pm 0.03 &	4.78 \pm 0.04\\
%\mathrm{HD}\,191408 &	0.88 &	1.8 &	4.72 &	4.62\\
\mathrm{HD}\, 23249 &	0.92 &	2.6 & \mathrm{FEROS} &	4.95 \pm 0.02 &	4.74 \pm 0.09\\
\mathrm{HD}\, 23249 & 0.92 &	2.6 & \mathrm{HARPS} & 4.02\pm0.02 & 3.78 \pm 0.05\\
\mathrm{HD}\, 82106 &	1.00 & 	3.1 & \mathrm{FEROS} &	5.04 \pm 0.02 &	4.74 \pm 0.11\\
\mathrm{HD}\, 32147 &	1.06 &	1.7 & \mathrm{FEROS} &	4.95 \pm 0.03 &	4.86 \pm 0.06\\
\mathrm{HD}\, 32147 & 1.06 &	1.7 & \mathrm{HARPS} & 3.94\pm0.02 & 3.84 \pm 0.03\\
\mathrm{HD}\,131977 &	1.11 &	2.6 & \mathrm{FEROS} &	5.06 \pm 0.02 &	4.85 \pm 0.09\\
\mathrm{HD}\, 81797 &	1.48 &	1.4^b & \mathrm{FEROS} &	5.26 \pm 0.06 &	5.20 \pm 0.08\\
\mathrm{HD}\, 81797 & 1.48 &	1.4^b & \mathrm{HARPS} & 4.43\pm0.02 & 4.37 \pm 0.02\\
%\mathrm{HD}\,152751 &	1.58 &	2.7 &	5.36 &	5.16\\
\hline
\hline
\end{array}
$$
The error for $v\sin i$ is 0.5\,km\,s$^{-1}$\\
References:\\
Valenti \& Fischer 2005\\
$^a$\,Reiners \& Schmitt~2003 \\
$^b$\,Gray \& Toner~1986
\label{tab_vsinikalib}
\end{table}

For FEROS data, the coefficients in Equation~\ref{eq_sigma0} are $a_0 = 0.641\pm0.001$, $a_1 = 0.043\pm0.003$, and for HARPS data the coefficients are $a_0 = 0.574\pm0.001$ and $a_1 = 0.032\pm0.002$.
\end{appendix}
%\listofobjects

\begin{thebibliography}{}
\bibitem{} Allain, S. 1998, A\&A, 333, 629
\bibitem{} Appenzeller, I. \& Mundt, R. 1989, A\&ARv, 1, 291
\bibitem{} Armitage, P.J. \& Clarke, C.J. 1996, MNRAS, 280, 458
\bibitem{} Attridge, J.M. \& Herbst, W. 1992, ApJ, 398, L61
\bibitem{} Baranne, A., Mayor, M., Poncet, J. L. 1979, Vistas astron., 23, 279
\bibitem{} Baranne, A., Queloz, D., Mayor, M., et al. 1996, A\&AS, 119, 373
\bibitem{} Benz, W., \& Mayor, M. 1984, A\&A, 138, 183
\bibitem{} Bodenheimer, P. 1989, NATO ASIC Proc.~290: Theory of Accretion Disks, 75
\bibitem{} Bodenheimer, P. 1995, ARA\&A, 33, 199
\bibitem{} Bouvier, J. 1993, The ESO Messenger, 71, 21
\bibitem{} Bouvier, J. 1995, MmSAI, 66, 341
\bibitem{} Bouwman, J., Lawson, W.~A., Dominik, C., et al. 2006, ApJ, 653, L57
\bibitem{} Brown, J.M., Blake, G.A., Dullemond, C.P., et al. 2007, ApJ, 664, L107
\bibitem{} Camenzind, M. 1990, Reviews in Modern Astronomy, 3, 234
\bibitem{} Carpenter, J.M., Bouwman, J., Silverstone, M.D., et al. 2008, ApJS, 179, 423
\bibitem{} Carpenter, J.M., Bouwman, J., Mamajek, E.E., et al. 2009, ApJS, 181, 197
\bibitem{} Chen, C.H., Patten, B.M., Werner, M.W., et al. 2005, ApJ, 634, 1372
\bibitem{} Cieza, L., \& Biliber, N. 2007, ApJ, 671, 605
\bibitem{} Fallscheer, C., \& Herbst, W. 2006, ApJ, 647, L155
\bibitem{} Fang, M., van Boeckel, R., Wang, W., et al. 2009, A\&A, 504, 461
\bibitem{} Flower, P.J. 1996, ApJ, 469, 355
\bibitem{} Gautier, T.N. III, Rebull, L.M., Stapelfeldt, K.R., et al. 2008, ApJ, 683, 813
\bibitem{} Ghez, A.M., McCarthy, D.W., Patience, J.L., \& Beck, T.L. 1997, ApJ, 481, 378
\bibitem{} Gray, D.F. \& Toner, C.G. 1986, ApJ, 310, 277
\bibitem{} Gray, D.F. 1992, The Observation and Analysis of Stellar Photospheres, 2nd Edition, Cambridge University Press
\bibitem{} Gray, R.O. \& Corbally, C.J. 1994, AJ, 107, 742
\bibitem{} Hartigan, P., Hartmann, L., Kenyon, S., Hewett, R., \& Stauffer, J. 1989, ApJS, 70, 899
\bibitem{} Hartmann, L. 2002, ApJ, 566, L29
\bibitem{} Herbst, W., Bailer-Jones, C. A. L., Mundt, R., et al. 2002, A\&A, 396, 513
\bibitem{} Herbst, W., Eisl\"offel, J., Mundt, R., \& Scholz, A. 2007, Protostars and Planets V, 297
\bibitem{} Herbst, W., Bailer-Jones, C.A.L. \& Mundt, R. 2001, ApJ, 554, 197
\bibitem{} Herbst, W. \& Mundt, R. 2005, ApJ, 633, 967
\bibitem{} Hirano, T., Suto, Y., Taruya, A., et al. 2010, ApJ, 709, 458
\bibitem{} Jayawardhana, R., Coffey, J., Scholz, A., et al. 2006, ApJ, 648, 1206
\bibitem{} Joergens, V. \& Guenther, E. 2001, A\&A, 379, L9
\bibitem{} Johns-Krull, C. M. \& Basri, G. 1997, ApJ, 474, 433
\bibitem{} Kaufer, A., Stahl, O., Tubbesing, S., et al. 1999, The ESO Messenger, 95, 8
\bibitem{} Kaufer, A., Stahl, O., Tubbesing, S., et al. 2000, SPIE, 4008, 459
\bibitem{} Kessler-Silacci, J., Augereau, J.C., Dullemond, C.P., et al. 2006, ApJ, 639, 275
\bibitem{} K\"ohler, R. 2001, AJ, 122, 3325
\bibitem{} K\"onigl, A. 1991, ApJ, 370, L39
\bibitem{} Kurucz, R. L. 1993, CD-ROM 13, ATLAS9 Stellar Atmosphere Programs and 2 km/s Grid (Cambridge: Smithsonian Astrophys. Obs.)
%\bibitem{} Lafreni\`ere, D., Jayawardhana, R., Brandeker, A., Ahmic, M. \& van Kerkwijk, M.~H. 2008, ApJj, 683, 844
\bibitem{} Lamm, M. H., Mundt, R., Bailer-Jones, C. A. L. \& Herbst, W. 2005, A\&A, 430, 1005
\bibitem{} Launhardt, R., Queloz, D., Henning, Th., et al. 2008, SPIE, 7013, 76
\bibitem{} Lommen, D., Wright, C.M., Maddison, S.T., et al. 2007, A\&A, 462, 211
\bibitem{} L\'opez-Santiago, J., Montes, D., Crespo-Chac\'on, I., \& Fern\'andez-Figueroa, M. J. 2006, ApJ, 643, 1160
\bibitem{} Low, F.J., Smith, P.S.. Werner, M., et al. 2005, ApJ, 631, 1170
\bibitem{} Makarov, V.V., \& Fabricius, C. 2001, A\&A, 368, 866
\bibitem{} Makidon, R.B., Rebull, L.M., Strom, S.E., et al. 2004, AJ, 127, 2228
\bibitem{} Mamajek, E.E., Meyer, M.R., \& Liebert, J. 2002, AJ, 124, 1670
\bibitem{} Mamajek, E.E., Meyer, M.R., Hinz, P.M., et al. 2004, ApJ, 612, 496
\bibitem{} Mathieu, R.D., Adams F.C., Latham, D.W. 1991, AJ, 101, 2184
\bibitem{} Matt, S. \& Pudritz, R.E. 2005, ApJ, 632, L135
\bibitem{} Matt, S. \& Pudritz, R.E. 2008, 14th Cambridge Workshop on Cool Stars, 384, 339
\bibitem{} Mayor, M., Pepe, F., Queloz, D., et al. 2003, The Eso Messenger, 114, 20
\bibitem{} Melo, C.H.F., Pasquini, L., \& De Medeiros, J.R. 2001, A\&A, 375, 851
\bibitem{} Muzerolle, J., Hartmann, L., \& Calvet, N. 1998, AJ, 116, 455
\bibitem{} Neuh\"auser, R., Guenther, E.W., Alves, J., et al. 2003, AN, 324, 535
\bibitem{} Nguyen, D.C., Jayawardhana, R., van Kerkwijk, M.H., et al. 2009, ApJ, 695, 1648
\bibitem{} Noyes, R. W., Hartmann, L. W., Baliunas, S. L., et al. 1984, ApJ, 279, 777
\bibitem{} Padgett, D., Koerner, D., Wahhaj, Z., et al. 2006, ApJ, 645, 1283
\bibitem{} Palla, F. 2002, in Physics of Star Formation in Galaxies, ed. Palla, F., Zinnecker, H., Maeder, A., \& Meynet, G. (Springer, Heidelberg), 9
\bibitem{} Pallavicini, R., Golub, L., Rosner, R., et al. 1981, ApJ, 248, 279
%\bibitem{} Queloz, D., Allain, S., Mermilliod, J.-C., Bouvier, J., \& Mayor, M. 1998, A\&A, 335, 183
\bibitem{} Rebull, L.M., Wolff, S.C., \& Strom, S.E. 2004, AJ, 127, 1029
\bibitem{} Rebull, L.M., Stauffer, J. R., Megeath, S. T., et al. 2006, ApJ, 646, 297
\bibitem{} Rebull, L.M., Stapelfeldt, K. R., Werner, M. W., et al. 2008, ApJ, 681, 1484
\bibitem{} Reiners, A. \& Schmitt, J.H.M.M. 2003, A\&A, 398, 647
\bibitem{} Rodr\'iguez-Ledesma, M.V., Mundt, R., \& Eisl\"offel, J. 2009, A\&A, 502, 883
\bibitem{} Saar, S.H. \& Donahue, R.A. 1997, ApJ, 485, 319
\bibitem{} Santos, N.C., Mayor, M., Naef, D., et al. 2002, A\&A, 392, 215
\bibitem{} Setiawan, J., Weise, P., Henning, Th., et al. 2007, ApJ, 660, 145
\bibitem{} Setiawan, J., Weise, P., Henning, Th., et al. 2008, Proceedings of Precision Spectroscopy in Astrophysics, ed. Santos, N.C. et al., 201
\bibitem{} Skumanich 1972, ApJ, 171, 565
\bibitem{} Soderblom, D.R., Jones, B.F., Balachandran, S., et al. 1993, AJ, 106, 1059
\bibitem{} Soderblom, D.R., King, J.R., Hanson, R.B., et al. 1998, ApJ, 504, 192
\bibitem{} Stassun, K.G., Mathieu, R.D., Mazeh, T. \& Vrba, F.J. 199, AJ, 117, 2941
\bibitem{} Torres, C.A.O., da Silva, L., Quast, G.R., de la Reza, R., \& Jilinski, E. 2000, AJ, 120, 1410
\bibitem{} Torres, C.A.O., Quast, G.R., de la Reza, R., da Silva, L., \& Melo, C.H.F. 2001, ASPC, 244, 43
\bibitem{} Torres, C.A.O., Quast, G.R., de la Reza, R., et al. 2006, A\&A, 460, 695
\bibitem{} Valenti, J.A. \& Fischer, D.A. 2005, ApJS, 159, 141
\bibitem{} White, R.J. \& Basri, G. 2003, ApJ, 582, 1109
\bibitem{} Wichmann, R., Schmitt, J.H.M.M., \& Hubrig, S. 2003, A\&A, 399, 983
\bibitem{} Weise, P. 2007, Diploma thesis, University of Heidelberg
\bibitem{} Wolff, S.C., Strom, S.E., \& Hillenbrand, L.A. 2004, ApJ, 601, 979
%\bibitem{} Yuan, F. 1998, Computer Aided Engineering, Nr. 3
\bibitem{} Zuckerman, B. \& Song, I. 2004, ARA\&A, 42, 685
\end{thebibliography}
\end{document}